\documentclass[nofootinbib,prd,twocolumn,showpacs,showkeys,preprintnumbers]{revtex4-1}
\usepackage{hyperref,amssymb,amsmath,mathrsfs,bm,graphicx}
\usepackage{multirow}
\usepackage{array}
\usepackage{dblfloatfix}
\usepackage{dblfloatfix}
\usepackage{placeins}
\usepackage{float}
\usepackage[dvipsnames]{xcolor}
\begin{document}

\title {Construction of a traversable wormhole from a suitable  embedding function.}

\author{A. Rueda }
\affiliation{Departamento de F\'isica, Colegio de Ciencias e Ingenier\'ia, Universidad San Francisco de Quito,  Quito 170901, Ecuador.\\}
\author{R. \'Avalos}
\affiliation{Departamento de F\'isica, Colegio de Ciencias e Ingenier\'ia, Universidad San Francisco de Quito,  Quito 170901, Ecuador.\\}
\author{E. Contreras}
\email{econtreras@usfq.edu.ec}
\affiliation{Departamento de F\'isica, Colegio de Ciencias e Ingenier\'ia, Universidad San Francisco de Quito,  Quito 170901, Ecuador.\\}

\begin{abstract}
In this work, we construct a traversable wormhole by providing a suitable embedding function ensuring the fulfilling of the flaring--out condition. The solution contains free parameters that are reduced through the study of the acceptable conditions of a traversable wormhole. We compute both the quantifier of exotic matter and the quasi--normal modes through the $13^{th}$ order WKB as a function of the remaining free parameters. We obtain that the wormhole geometry can be sustained by a finite amount of exotic matter and seems to be stable under scalar perturbations.  
\end{abstract}
\maketitle
\section{Introduction}
Since the seminal work by Morris and Thorne \cite{Morris:1988cz}, the study of traversable wormholes has remain as an attractive research line for many years given the intriguing features it encodes \cite{Morris:1988tu,Alcubierre:2017pqm,Visser:1995cc,Lobo:2005us,Garattini:2019ivd,Stuchlik:2021guq,Bronnikov:2021ods,Blazquez-Salcedo:2021udn,Churilova:2021tgn,Konoplya:2021hsm,Tello-Ortiz:2021kxg,Bambi:2021qfo,Capozziello:2020zbx,Blazquez-Salcedo:2020czn,Berry:2020tky,Maldacena:2020sxe,Garattini:2020kqb}.
However, the detection of gravitational waves by LIGO/Virgo from the merger of binary black holes \cite{LIGOScientific:2016aoc,LIGOScientific:2016sjg,LIGOScientific:2017bnn} has driven the attention
on the possibility of considering wormholes as black holes mimickers \cite{Bueno:2017hyj,Cardoso:2016oxy,Konoplya:2016hmd}. To be more precise, as the ringdown phase of black hole mergers is dominated  by the quasinormal modes of the final object, it has been claimed that wormholes can mimic black holes based on the similitude of their quasinormal modes spectrum.

From a technical point of view, any traversable wormhole could be constructed by providing its geometry in terms of free parameters which should be constrained based on the acceptability conditions it must satisfy: i) the existence of a throat connecting two asymptotically flat regions, ii) small tidal forces bearable by a human being iii) finite proper time to traverse the throat, among others. However, the fulfilling of all the requirements is not always possible. For example, the solution should not be asymptotically flat or should require an infinite amount of exotic matter supporting it. 
It should be emphasized that the above mentioned requirements are not universal
but sufficient for a wormhole to be a physically viable and suitable
for interstellar travel by human beings. For example, we can construct asymptotically AdS wormholes which are suitable for interstellar travels. \cite{Blazquez-Salcedo:2020nsa}. Nevertheless,
in this work, it is our main goal to construct an asymptotically flat traversable wormhole supported by a finite amount of exotic matter by assuming a general embedding function. Besides, we explore its stability thorough its response to scalar perturbation.  

The response of a wormhole to perturbations is dominated by damped oscillations called quasi--normal modes. The computation of the QNM modes can be performed through a variety of methods (for an incomplete list see \cite{Konoplya:2022tvv,Churilova:2021nnc,Konoplya:2020jgt,Konoplya:2020bxa,Konoplya:2019nzp,Rincon:2021gwd,Panotopoulos:2020mii,Rincon:2020cos,Rincon:2020iwy,Rincon:2020pne,Xiong:2021cth,Zhang:2021bdr,Panotopoulos:2019gtn,Lee:2020iau,Churilova:2019qph,Oliveira:2018oha,Panotopoulos:2017hns} and references therein, for example). However, in this work, we shall use the recently developed  WKB approximation to the $13^{th}$ order which has brought the attention of the community \cite{Konoplya:2019hlu}. It should be emphasized that, for the application of the method in the context of traversable wormholes, a bell--shape potential must be ensured. In this work, we study the QNM for the model after providing the suitable sets of parameters that ensure a bell--shaped potential.

This work is organized as follows. In section \ref{TW} we review the mains aspects related to traversable wormholes. Next, in section \ref{model} we propose the embedding function, obtain the shape of the wormhole and analyse the quantifier of the exotic matter. In section \ref{QNM} we implement the $13^{th}$ order WKB approximation to compute and interpret the quasinormal modes associated to the scalar perturbations of the wormhole. Finally, in the last section we conclude the work.

\section{Traversable wormholes}\label{TW}
Let us consider the spherically symmetric line element
\begin{eqnarray}\label{metric}
ds^{2}=-e^{2\phi} dt^2 +d r^{2}/(1-b/r)+r^{2}(d\theta^{2}+\sin^{2}\theta d\phi^{2}),\nonumber\\
\end{eqnarray}
with $\phi=\phi(r)$ and $b=b(r)$ the redshift and shape functions respectively. 
Assuming that (\ref{metric}) is a solution of Einstein's equations
\begin{eqnarray}\label{EFE}
R_{\mu\nu}-\frac{1}{2}g_{\mu\nu}R=\kappa T_{\mu\nu},
\end{eqnarray}
with $\kappa=8\pi G/c^{4}$ \footnote{In this work we shall assume $c=G=1$.}, sourced by 
$T^{\mu}_{\nu}=diag(-\rho,p_{r},p_{t},p_{t})$
we arrive at
\begin{eqnarray}
\rho&=&\frac{1}{8\pi}\frac{b'}{r^{2}}\label{rho}\\
p_{r}&=&-\frac{1}{8\pi}\left[\frac{b}{r^{3}}-2\left(1-\frac{b}{r}\right)\frac{\phi'}{r}\right]\label{pr}\\
p_{t}&=&\frac{1}{8\pi}\left(1-\frac{b}{r}\right)
\bigg[\phi''+(\phi')^{2}-\frac{b'r-b}{2r^{2}(1-b/r)}\phi'\nonumber\label{pt}\\
&&-\frac{b' r-b}{2r^{3}(1-b/r)}+\frac{\phi'}{r}\bigg].
\end{eqnarray}

In what follows we shall describe the main aspects of a traversable wormhole by its embedding in  the three dimensional Euclidean space. First, note that as our solution is spherically symmetric, 
we can consider $\theta=\pi/2$ without loss of generality. Now, considering a fixed time, $t=constant$, the line element reads
\begin{eqnarray}\label{emb1}
ds^{2}=\frac{dr^{2}}{1-b/r}+r^{2}d\phi^{2}.
\end{eqnarray}
The surface described by (\ref{emb1}) can be embedded in $\mathbf{R}^{3}$ where the metric in cylindrical coordinates $(r,\phi,z)$ reads
\begin{eqnarray}
ds^{2}=dz^{2}+dr^{2}+r^{2}d\phi^{2}.
\end{eqnarray}
Next, as $z$ is a function of the radial coordinate we have
\begin{eqnarray}
dz=\frac{dz}{dr}dr,
\end{eqnarray}
from where
\begin{eqnarray}\label{emb2}
ds^{2}=\left[1+\left(\frac{dz}{dr}\right)^{2}\right]dr^{2}
+r^{2}d\phi^{2}.
\end{eqnarray}
Finally, from (\ref{emb1}) and (\ref{emb2}) we obtain
\begin{eqnarray}\label{emb3}
\frac{dz}{dr}=\pm 
\left(\frac{r}{b}-1\right)^{-1/2},
\end{eqnarray}
where is clear that $b>0$ for $r\in[r_{0},\infty)$.
At this point some comments are in order. First, the wormhole geometry must be endowed with minimum radius which leads to $dz/dr\to\infty$ as $r\to b_{0}$ (that occurs when $b=r$). Accordingly, the
existence of a minimum
radius requires that at $r=b_{0}$ the shape function must be $b=b_{0}$. Second, we demand that the solution is asymptotically flat which implies both, $b/r\to0$ (from where $dz/dr\to0$) and $\phi\to0$ as $r\to\infty$.  Third, as
as the conditions 
\begin{eqnarray}
&&\lim\limits_{r\to b_{0}}\frac{dz}{dr}\to\infty\label{c1}\\
&&\lim\limits_{r\to\infty}\frac{dz}{dr}=0\label{c2},
\end{eqnarray} 
must be satisfied, the smoothness of the geometry is ensured whenever the embedding surface flares out at or near the throat, namely  
\begin{eqnarray}
\frac{d^{2}r}{dz^{2}}>0,
\end{eqnarray}
from where
\begin{eqnarray}\label{foc}
\frac{b-b'r}{2b^{2}}>0,
\end{eqnarray}
which corresponds to the 
flaring--out condition. 

It is worth mentioning that, the flaring out condition (\ref{foc}) leads to the violation of the null energy condition (NEC) as we shall see in what follows. Let us define the quantity
\begin{eqnarray}
\xi=-\frac{p_{r}+\rho}{|\rho|}=\frac{b/r-b'-2(r-b)\phi'}{|b'|},
\end{eqnarray}
which can be written as
\begin{eqnarray}
\xi=\frac{2b^{2}}{r|b'|}\frac{d^{2}r}{dz^{2}}
-2(r-b)\frac{\phi'}{|b'|}
\end{eqnarray}
Now, as $(r-b)\to0$ at the throat, we have 
\begin{eqnarray}
\xi=\frac{2b^{2}}{r|b'|}\frac{d^{2}r}{dz^{2}}>0
\end{eqnarray}
so that
\begin{eqnarray}
\xi=-\frac{p_{r}+\rho}{|\rho|}>0.
\end{eqnarray}
Note that if $\rho>0$ the above condition implies $p_{r}<0$ which entails that $T^{1}_{1}$ should be interpreted as a tension. Furthermore, if we define $\tau=-p_{r}$ the flaring out condition leads to
\begin{eqnarray}
\tau-\rho>0,
\end{eqnarray}
which implies that, for this exotic matter, the throat tension must be greater than the total energy density which violates the NEC, as we stated before. Although there is not evidence of exotic matter in the universe, we can 
minimize the amount required to construct a traversable wormhole by demanding that the quantifier \cite{Visser:2003yf}
\begin{eqnarray}\label{visserQ}
I=\int dV (\rho+p_{r})=-\int\limits_{r_{0}}^{\infty}
(1-b')\left[\ln \left(\frac{e^{2\phi}}{1-b/r}\right)\right] dr \nonumber\\
\end{eqnarray}
is finite.\\



From a technical point of view, the construction of traversable wormholes require solving the system (\ref{rho})-(\ref{pt}), namely three equations with 
five unknowns.  The strategy should be either supplying the metric functions that satisfy the geometric constraints listed above or giving one of the metrics and an auxiliary condition, namely an equation of state or a metric constraint. In this work we follow an alternative route which consists of proposing a suitable embedding function.

\section{A wormhole model}\label{model}
\begin{figure*}[ht!]
\centering
\includegraphics[width=0.35\textwidth]{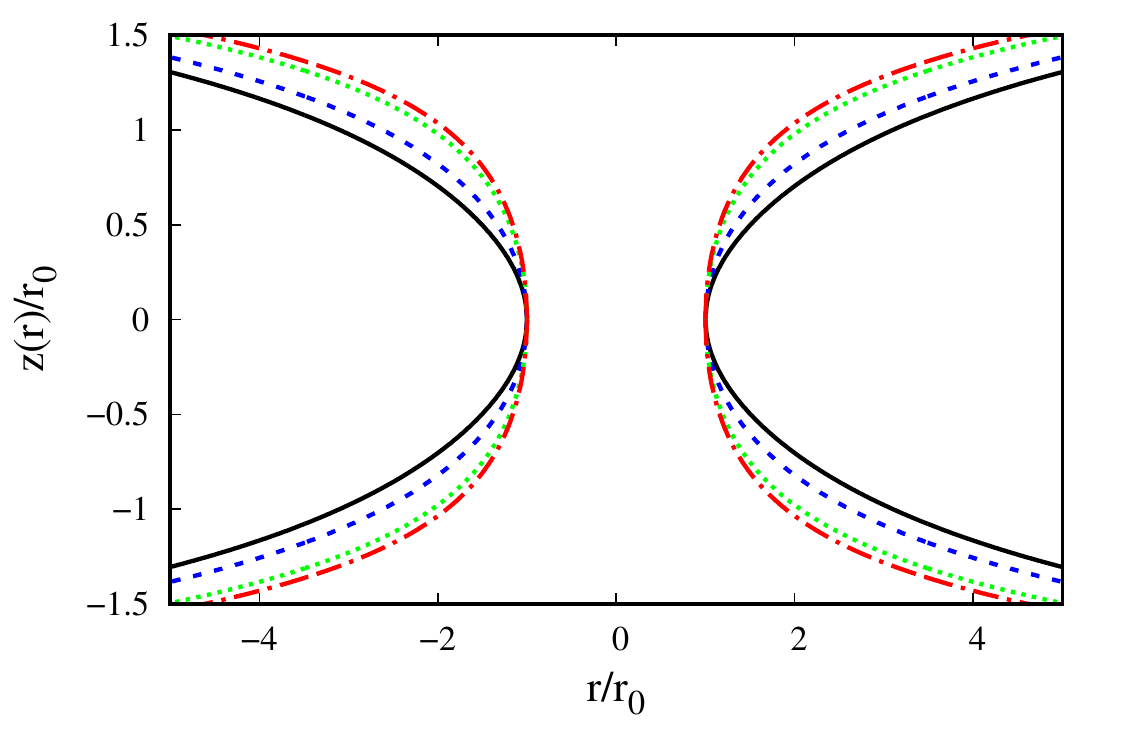}  \
\includegraphics[width=0.35\textwidth]{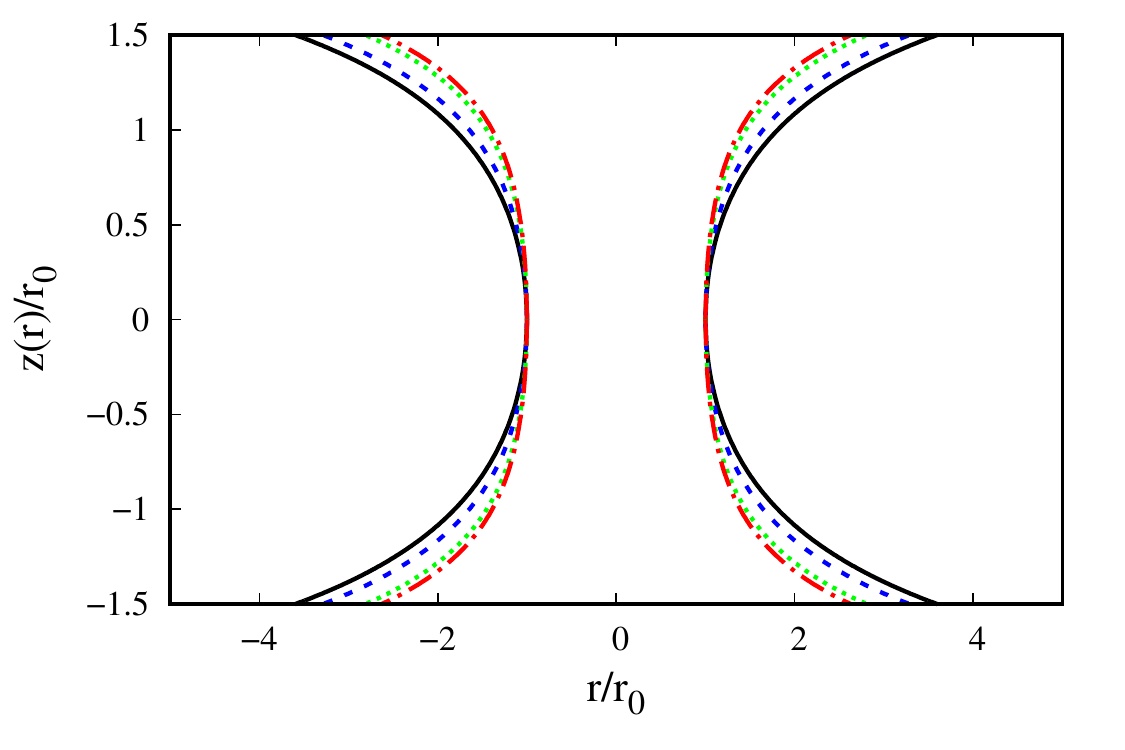}  \
\includegraphics[width=0.35\textwidth]{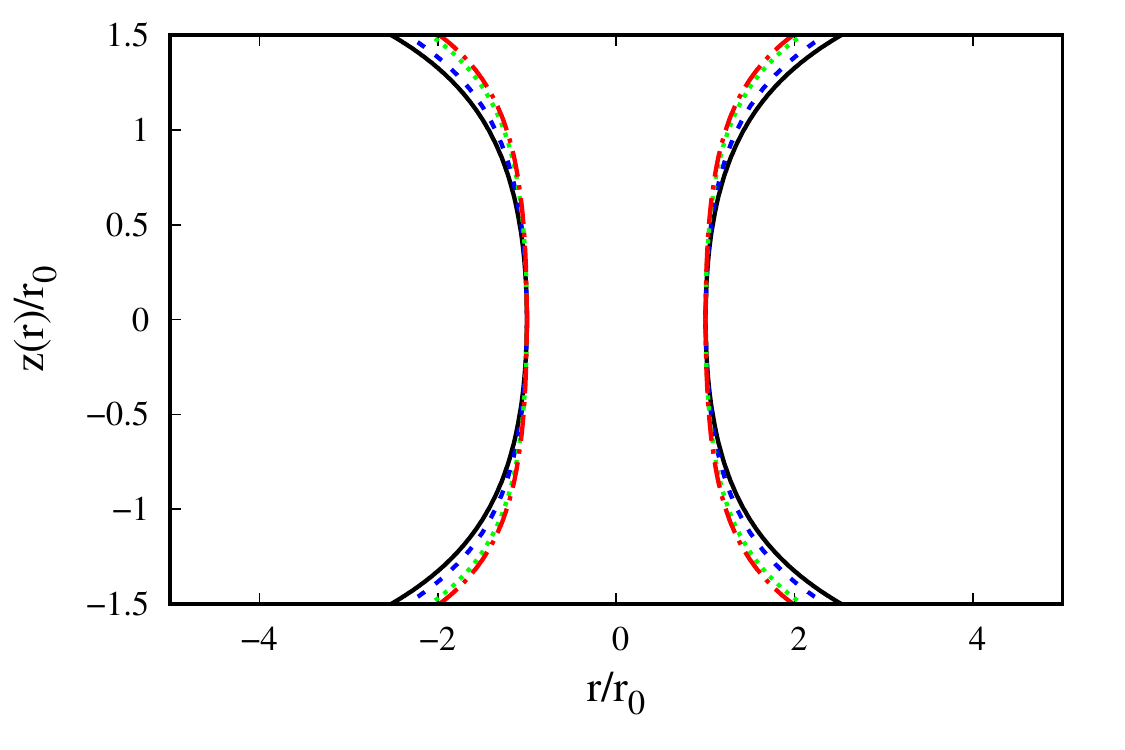}  \
\includegraphics[width=0.35\textwidth]{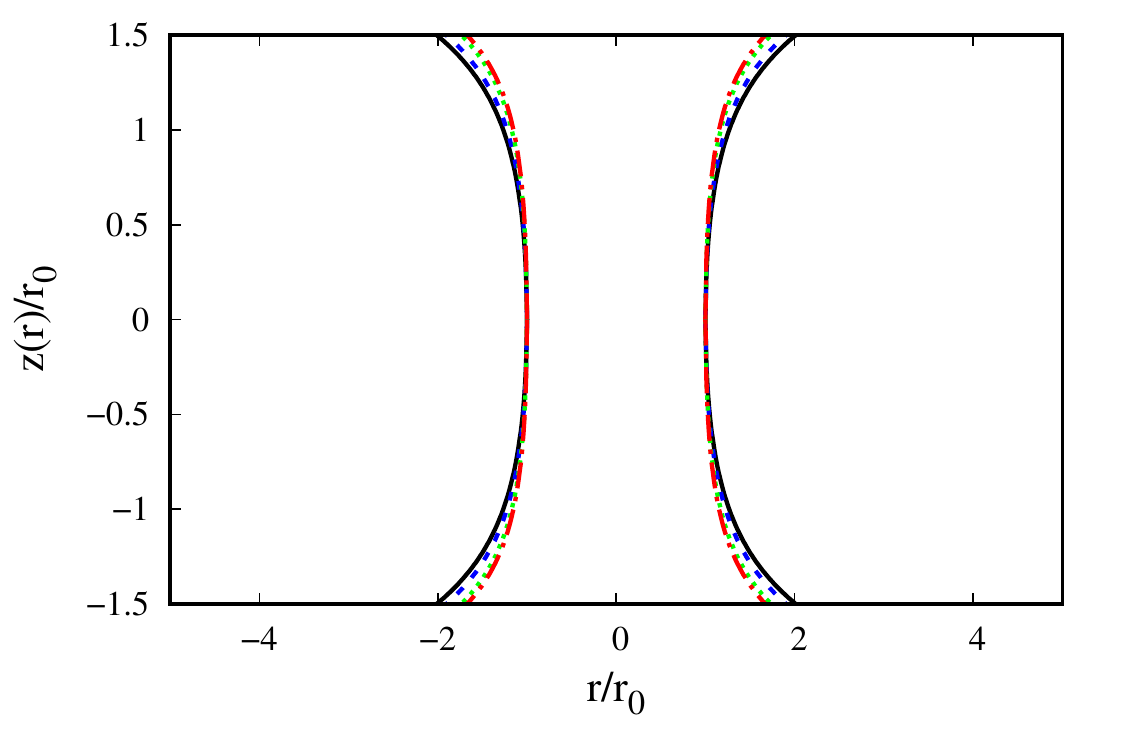}  
\caption{\label{plotembedding}
Embedding diagrams for $c = 0.4$ (first row, left panel), $0.8$ (first row, right panel), $1.2$ (second row, first panel), $1.6$ (second row, right panel) and $d = 0.2$ (black), $d =
0.6$ (blue), $d = 1.4$ (green), $d = 1.8$ (red)}
\end{figure*}

In this section, we construct a traversable wormhole geometry by providing a general embedding function with the aim to integrate Eq. (\ref{emb3}) and obtain the shape function $b$. Note that, although Eq. (\ref{emb3}) can  always be inverted numerically, in this work we look for analytical solutions so we propose
\begin{equation}\label{z}
  z (r)=\sqrt {\log\left (a + \left (\frac {c r} {r_0} + d \right)^2 \right)}  
\end{equation}
where $a$, $c$ and $d$ are free parameters. The free parameters can be constrained by imposing both the existence of a throat (a minimum radius $r_{0}$) and the flaring out condition given by Eqs. (\ref{c1}) and (\ref{c2}) which lead to
\begin{equation}
    a = 1 - c^2 - 2 c d - d^2
\end{equation}
In figure \ref{plotembedding} we show the embedding function for different values of the parameters involved. Note that the profiles flare out
slowly as both $c$ and $d$ increases.

\begin{figure*}[hbt!]
\centering
\includegraphics[width=0.35\textwidth]{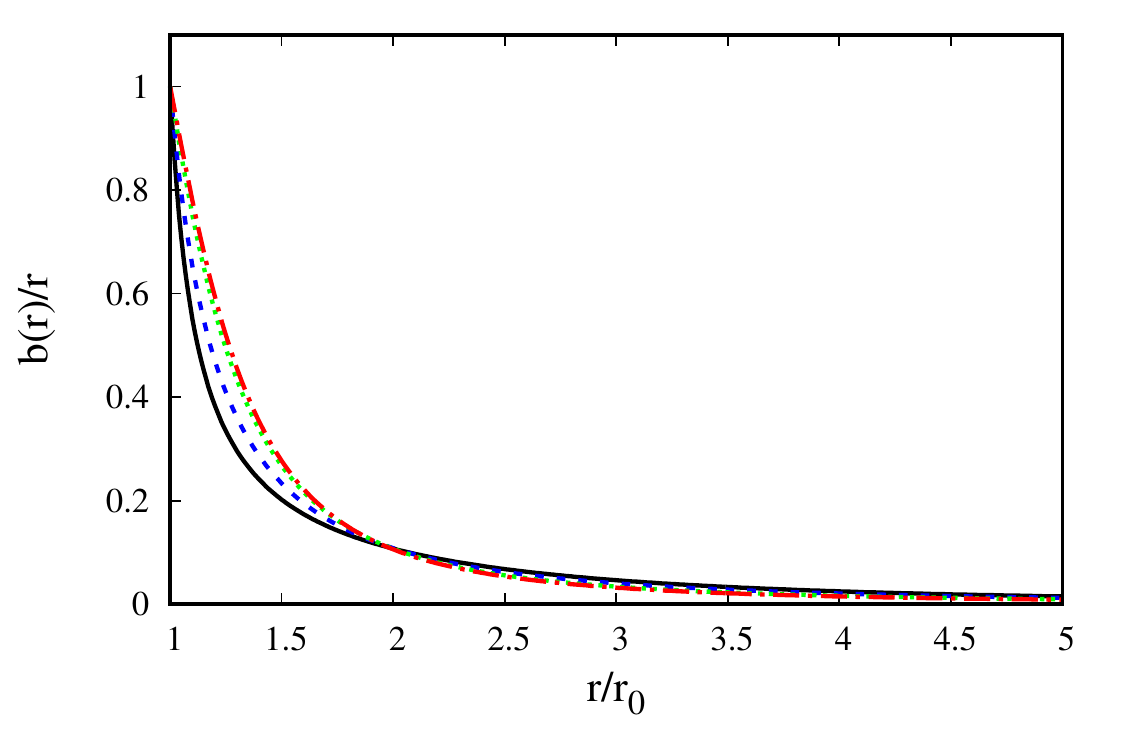}  \
\includegraphics[width=0.35\textwidth]{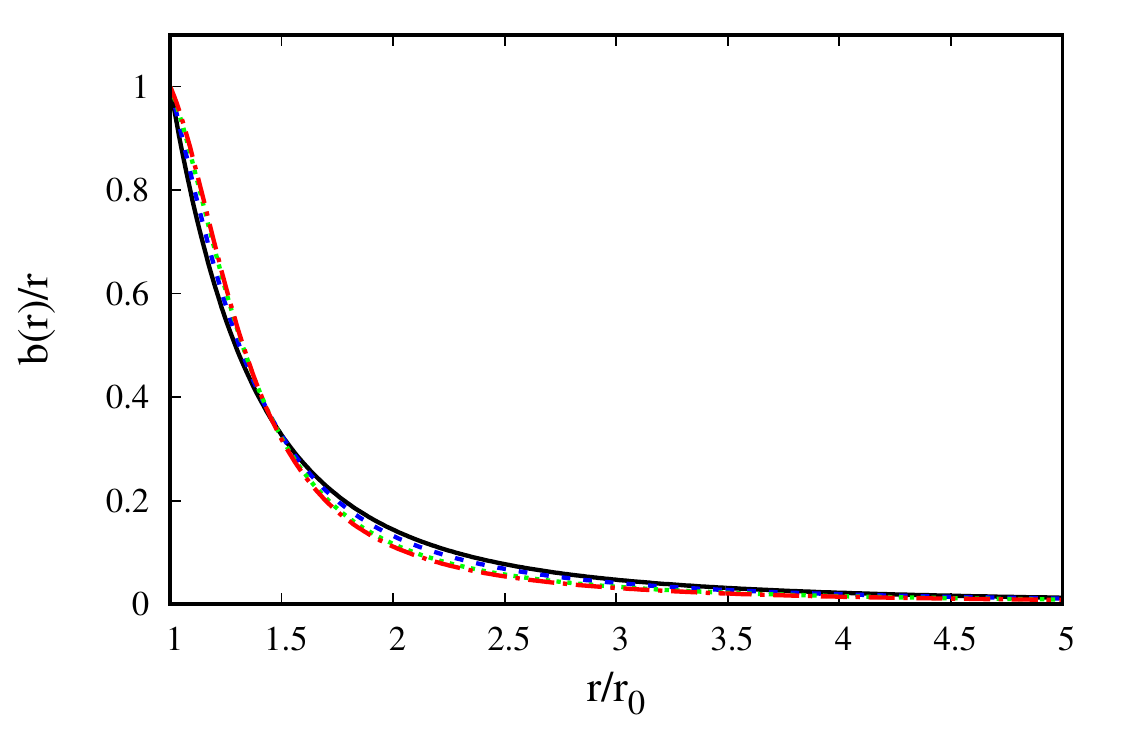}  \
\includegraphics[width=0.35\textwidth]{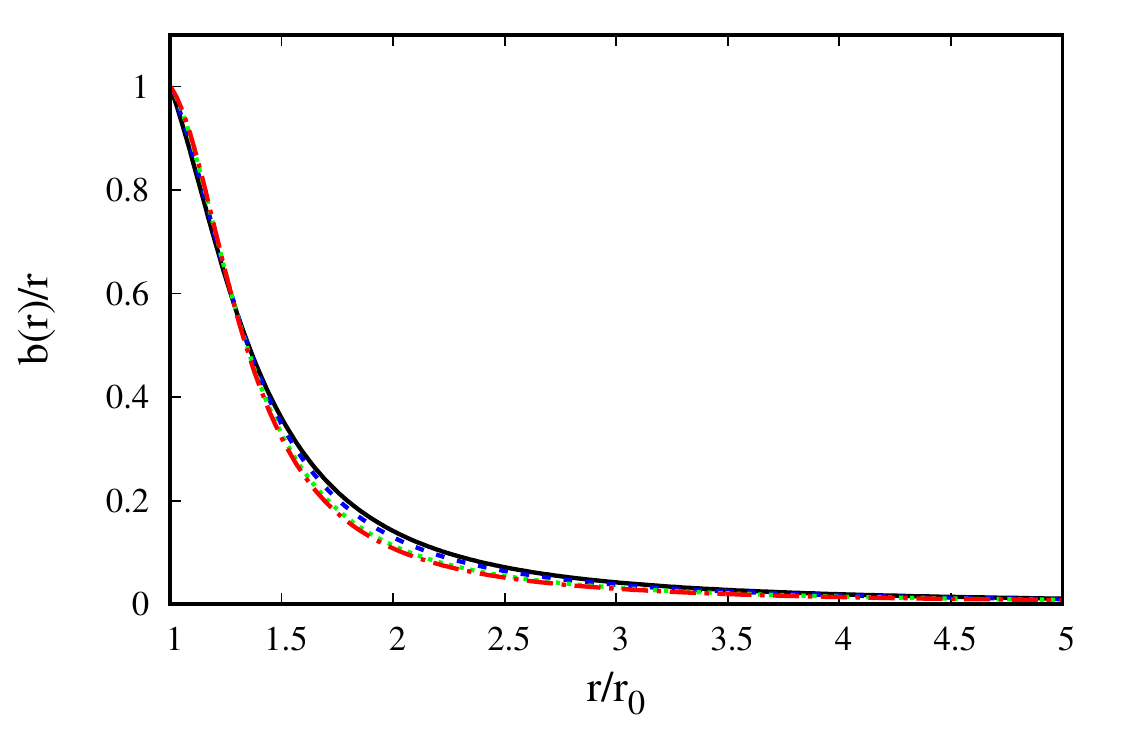}  \
\includegraphics[width=0.35\textwidth]{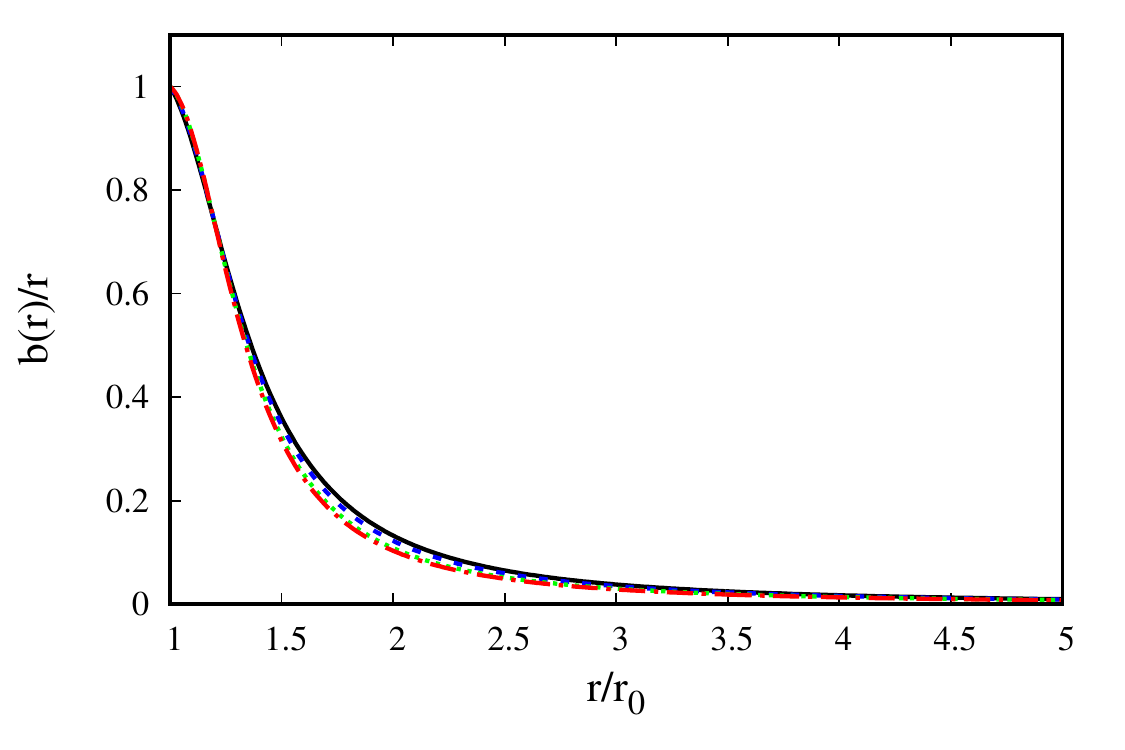}
\caption{\label{plotbr}
Plots of $b/r$ as a function of $r/r_0$ for $c = 0.4$ (first row, left panel), $0.8$ (first row, right panel), $1.2$ (second row, left panel), $1.6$ (second row, right panel) and d = 0.2 (black), $d =
0.6$ (blue), $d = 1.4$ (green), $d = 1.8$ (red)}
\end{figure*}

The shape function is obtained by replacing (\ref{z}) in (\ref{emb3}). As a result we obtain
\begin{equation}
       b(r)=\frac{c^2\zeta^2 r}{c^2\zeta^2+r_0^2\left( a +\zeta^2 \right)^2\log\left( a+\zeta^2\right)}
\end{equation}
where
\begin{eqnarray}
\zeta=\left( \frac {c r} {r_0} +d \right).
\end{eqnarray}
Note that the conditions $z(r_0)=0$ and $b(r_0)=r_0$ hold, as expected. In figure \ref{plotbr} we show the embedding function and the shape function for different values of the parameters involved. Note that, the solution is asymptotically flat as required.

\begin{figure*}[hbt!] 
\centering
\includegraphics[width=0.35\textwidth]{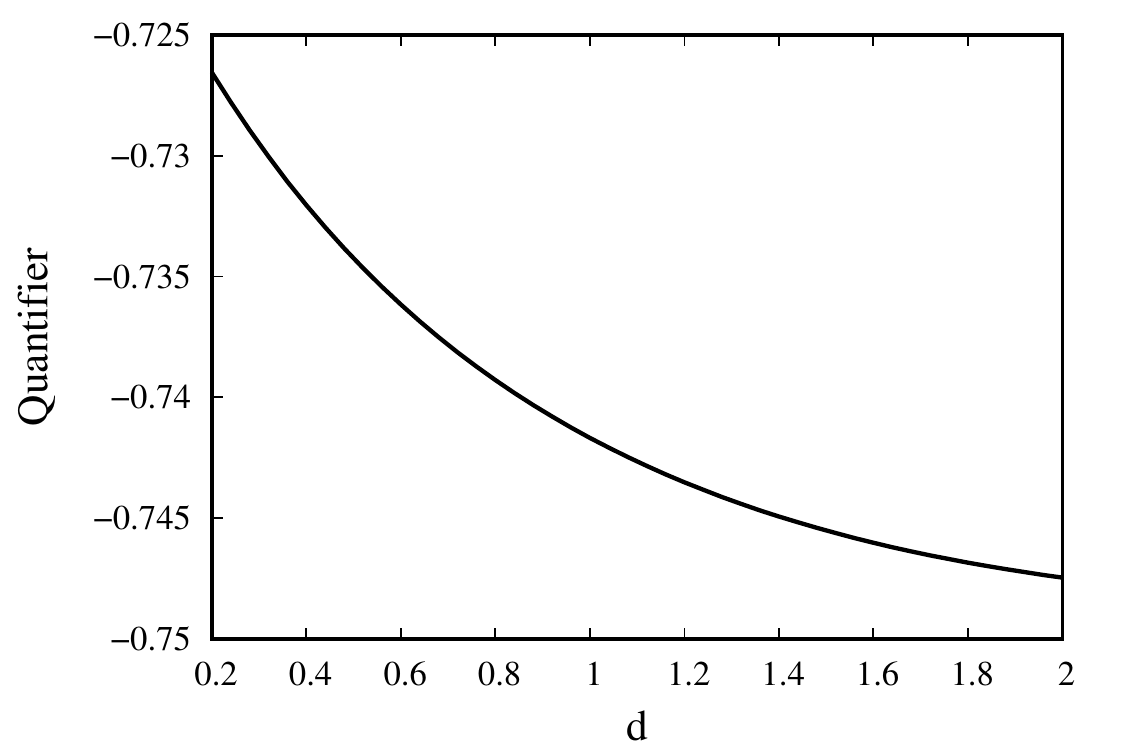}  \
\includegraphics[width=0.35\textwidth]{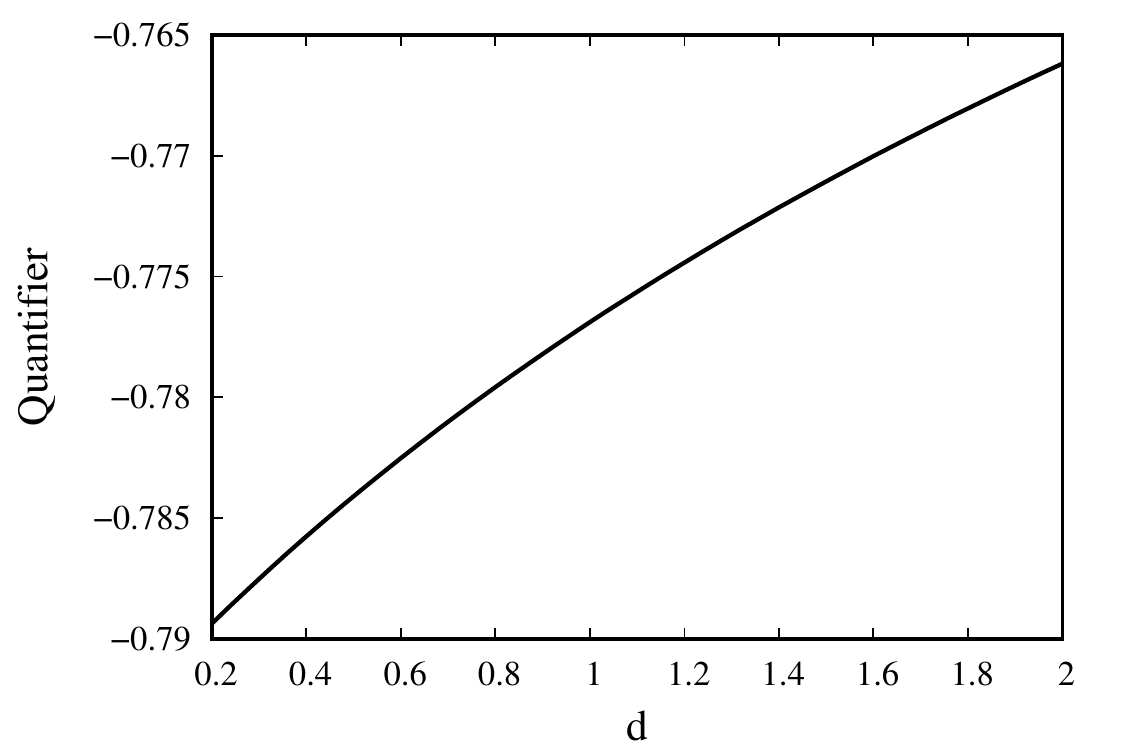}  \
\includegraphics[width=0.35\textwidth]{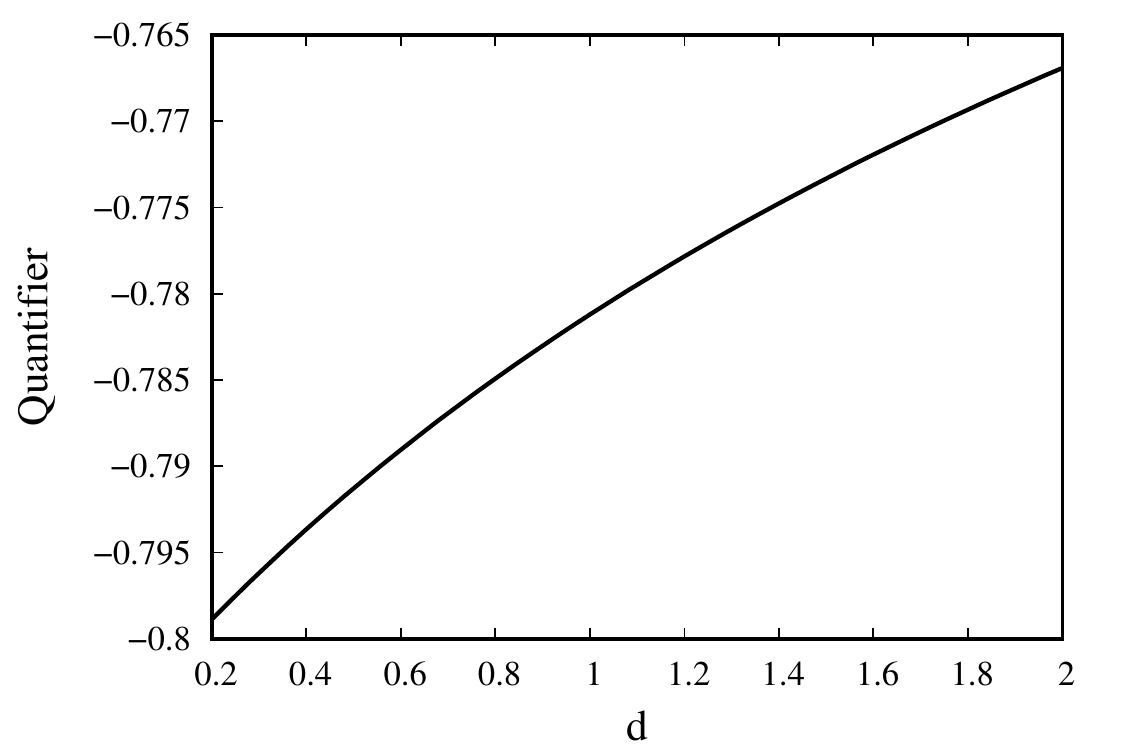}  
\includegraphics[width=0.35\textwidth]{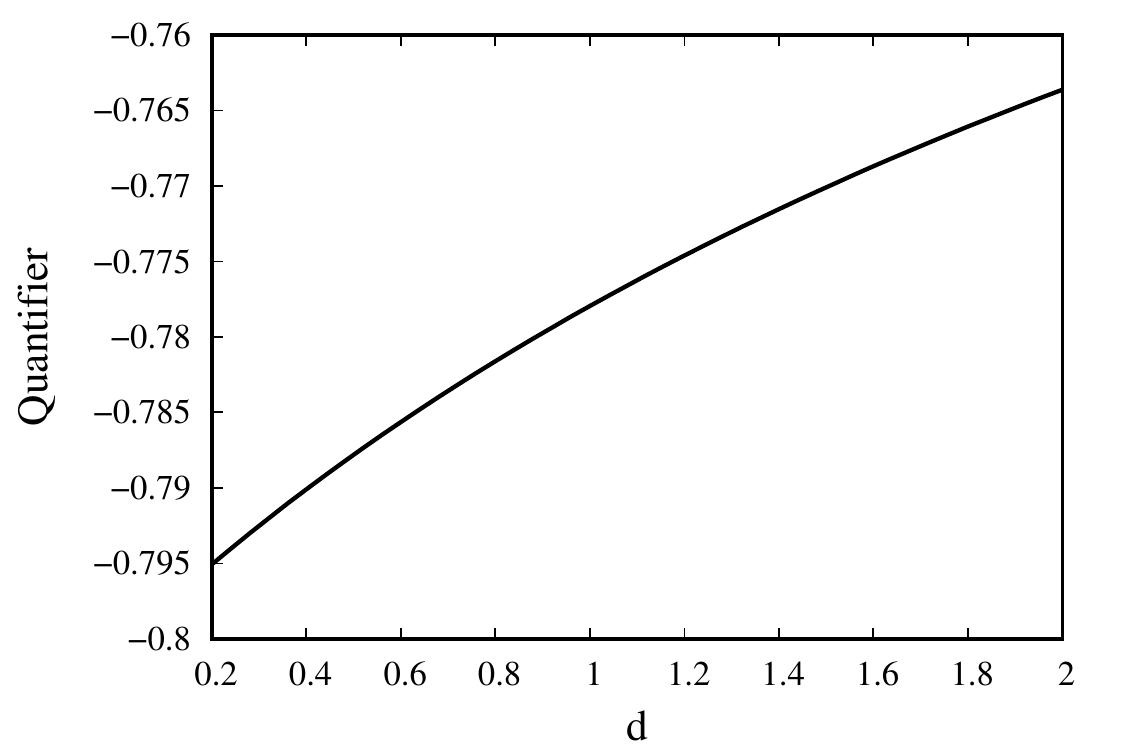}  \

\caption{\label{quanti}
Quantifier as a function of $d$ for $c = 0.4$ (first row, left panel), $0.8$, (first row, right panel) $1.2$ (second row, left panel), $1.6$ (second row, right panel).}
\end{figure*}

In order to specify the wormhole metric completely, we must propose a suitable redshift function. In this work we shall take the simplest choice, namely $\phi=0$, so the solution has a vanishing radial tidal force. At this point, the matter sector can be completely specified but the expressions of the density and pressure are too long to be shown here. Nevertheless, a more interesting issue is the analysis of the quantifier of the exotic matter which is finite for all the parameters under consideration as shown in Fig. (\ref{quanti}). Besides, 
for $c=0.4$, the amount of exotic matter required to sustain the wormhole increases as $d$ grows.
In contrast, for $c=\{0.8,1.2,1.6\}$ the decreasing of the amount of exotic matter is associated with an increasing of the free parameter $d$. 


\section{QNM by the WKB approximation}\label{QNM}
The perturbations of the TW can be carried out by adding test fields (scalar or vectorial) to the background or through perturbations of the space--time itself. However, independently of how it is performed, the equation governing the evolution of the perturbation can be reduced to a like--Schr\"{o}dinger equation given by
\begin{eqnarray} \label{schrodinger}
\bigg( \frac{d^2}{dr_*^2} +\omega ^2 -V(r_*) \bigg)\chi(r_*)=0,
\end{eqnarray}
where $r_*$ is the tortoise radial coordinates defines as
\begin{eqnarray}
r_*(r)=\int_{r_0}^r \frac{1}{\sqrt{1-b(r')/r'}}dr',
\end{eqnarray}
and $V(r)$ is an effective potential. Notices that the tortoise coordinate $r_*$ is defined in the interval $(-\infty, \infty)$ in such way that the spatial infinity at both sides of the wormhole corresponds to $r_*=\pm\infty$ and the wormhole throat is located at $r_*=0$. In this work, we will focus on the study of scalar perturbations, so the effective potential takes the form 
\begin{eqnarray} \label{veff}
V_L (r) = e^{2\phi} \bigg(\frac{L(L+1)}{r^2}-\frac{r b' -b}{2r^3} +\frac{\phi'}{r} \bigg(1-\frac{b}{r} \bigg) \bigg),
\end{eqnarray}
where $L$ is called the multipole number or fundamental tone. 
The solution of (\ref{schrodinger}) 
with the boundary conditions
\begin{eqnarray} \label{bc}
\chi (r_*) \sim C_{\pm}\exp (\mp i \omega r_*), \,\,\,\,\, r_*\rightarrow \pm \infty,
\end{eqnarray}
corresponding to purely out-going waves at infinity, are the QNM with frequency $\omega=Re(\omega)+iIm(\omega)$. The real part of of the QNM frequencies corresponds to the frequency of oscillation, while the imaginary part $Im(\omega)$ relates with the damping factor due to the loss of energy produced by the gravitational radiation. It is worth noticing that when $Im(\omega)>0$, the perturbation grows exponentially meaning an instability in the system. For the system to be stable, is required that $Im(\omega)<0$. Also consider that a complete perturbation will be a superposition of different tones $L$, which means that we need that every possible frequency satisfy $Im(\omega)<0$.
The QNM frequencies are usually obtained by numerical methods but, in this work, we shall implement the WKB approach taking advantage of the similarity of Eq. (\ref{schrodinger}) with the one-dimensional Schr\"{o}dinger equation with a potential barrier. This method was first used by Schutz and Will in Ref. \cite{Schutz:1985km} to study scattering around black holes and has been extended to higher orders around the top of the bell-shaped potential \cite{Churilova:2019qph}. Specifically, the $13^{th}$ order formula reads
\begin{eqnarray} \label{WKB}
i \frac{\omega ^2 -V_0}{\sqrt{-2V_0''}}-\sum_{j=2}^{13}\Lambda_j=n+\frac{1}{2},
\end{eqnarray}
being $V_0$ the maximum height of the potential and $V_0''$ its second derivative with respect to the tortoise coordinate evaluated at the radius where $V_{0}$ reaches a maximum which for a TW with a bell--shaped potential occurs at the throat ($r=r_0$, $r_*=0$). The higher order corrections are encoded in $\Lambda_j$ which depend on the value of the potential its derivatives evaluated at the maximum. The exact expressions for this corrections can be found in \cite{Konoplya:2019hlu}.

\begin{figure*}[hbt!]
\centering
\includegraphics[width=0.3\textwidth]{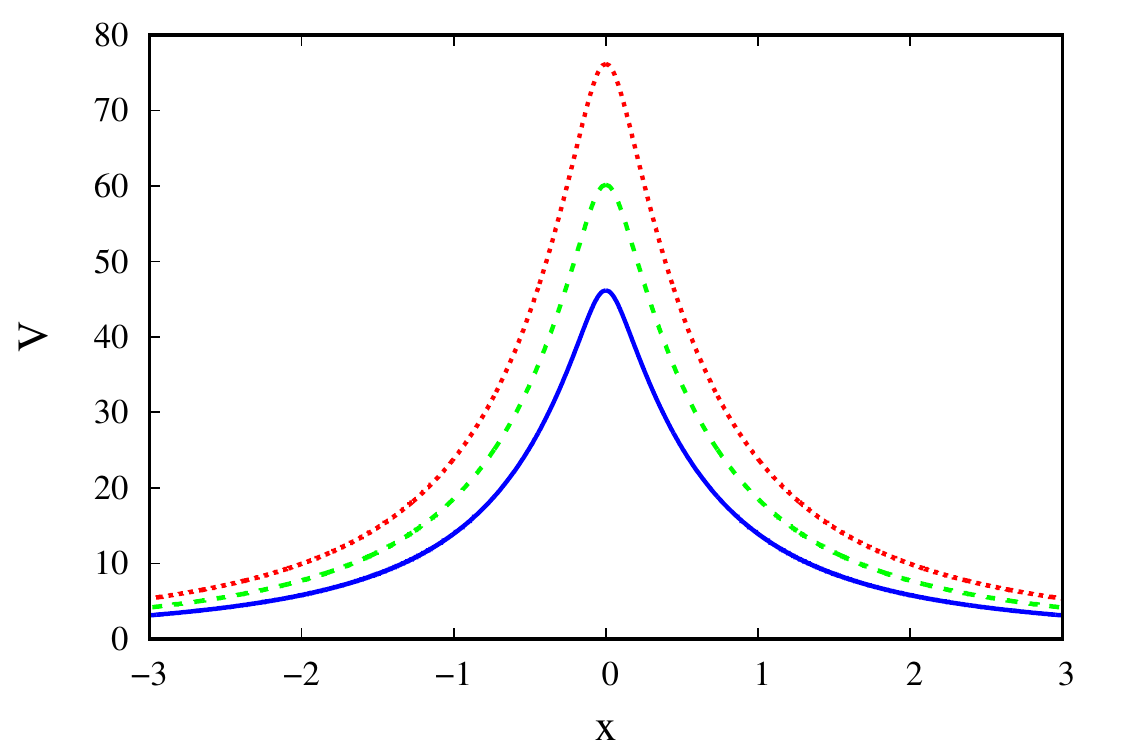}  \
\includegraphics[width=0.3\textwidth]{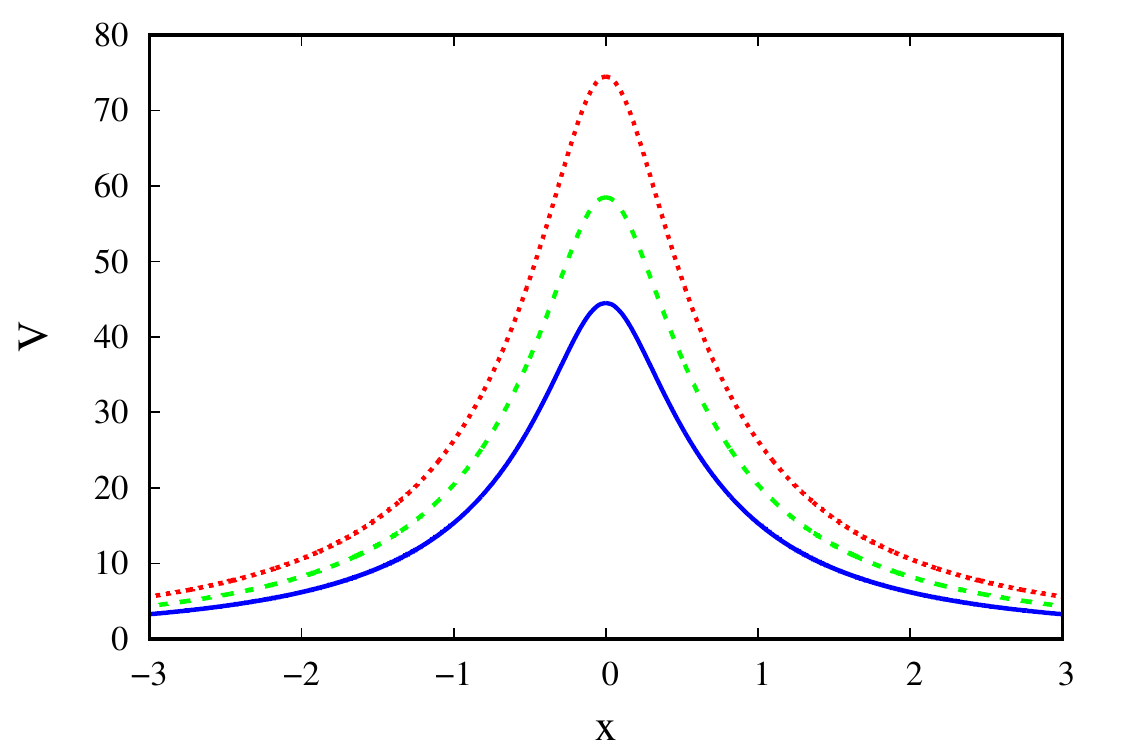}  \
\includegraphics[width=0.3\textwidth]{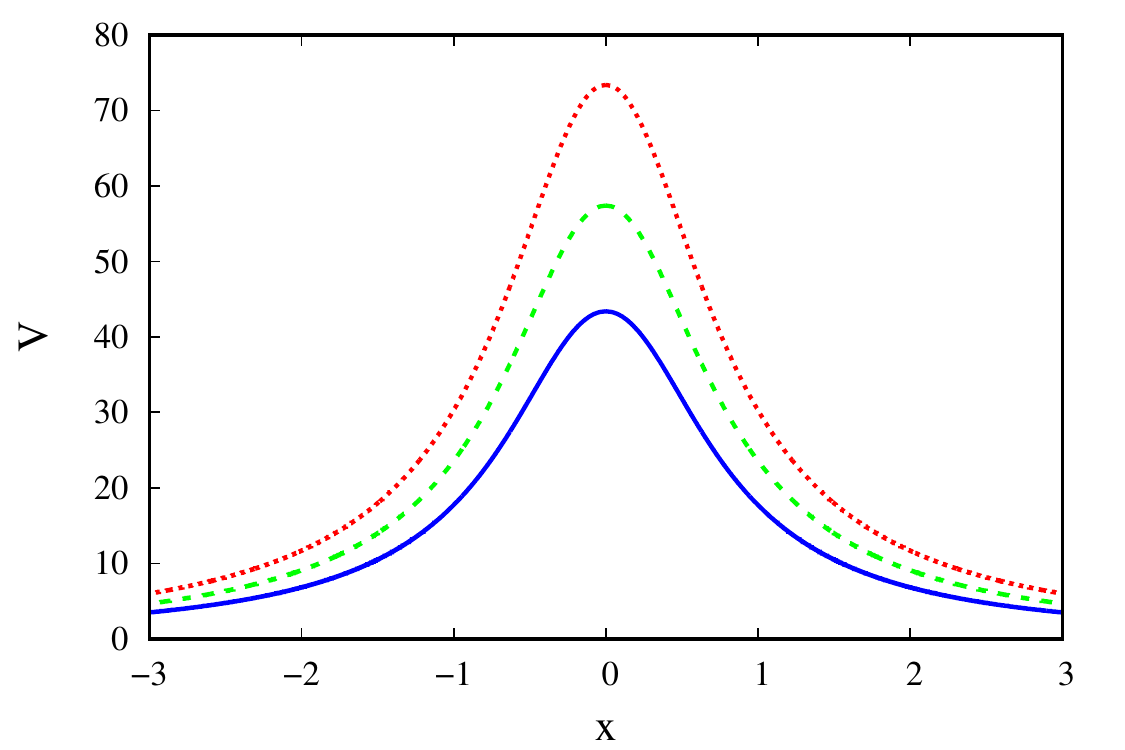}  
\includegraphics[width=0.3\textwidth]{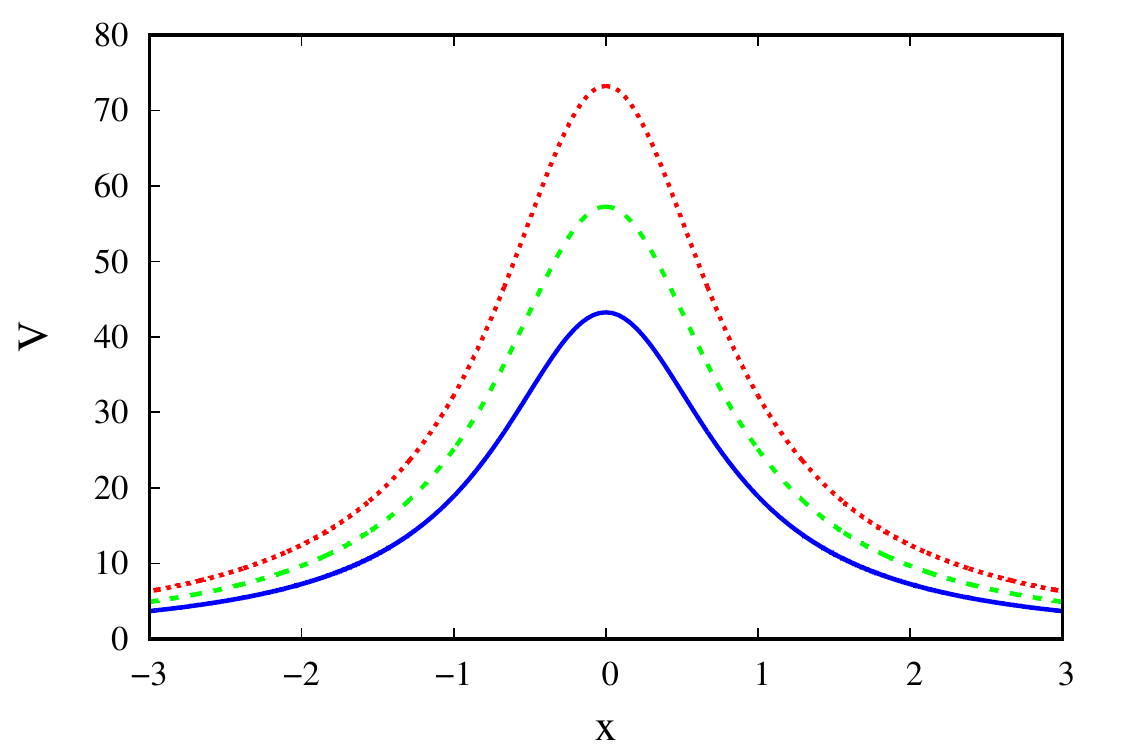}  \
\includegraphics[width=0.3\textwidth]{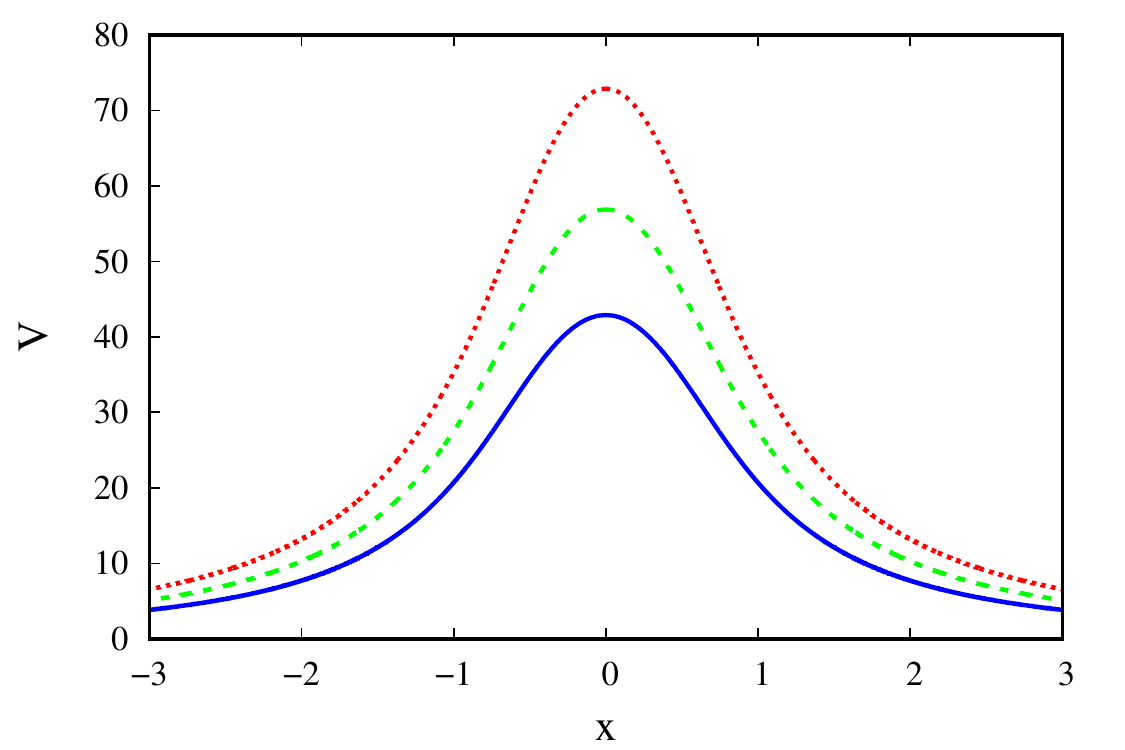}  \
\includegraphics[width=0.3\textwidth]{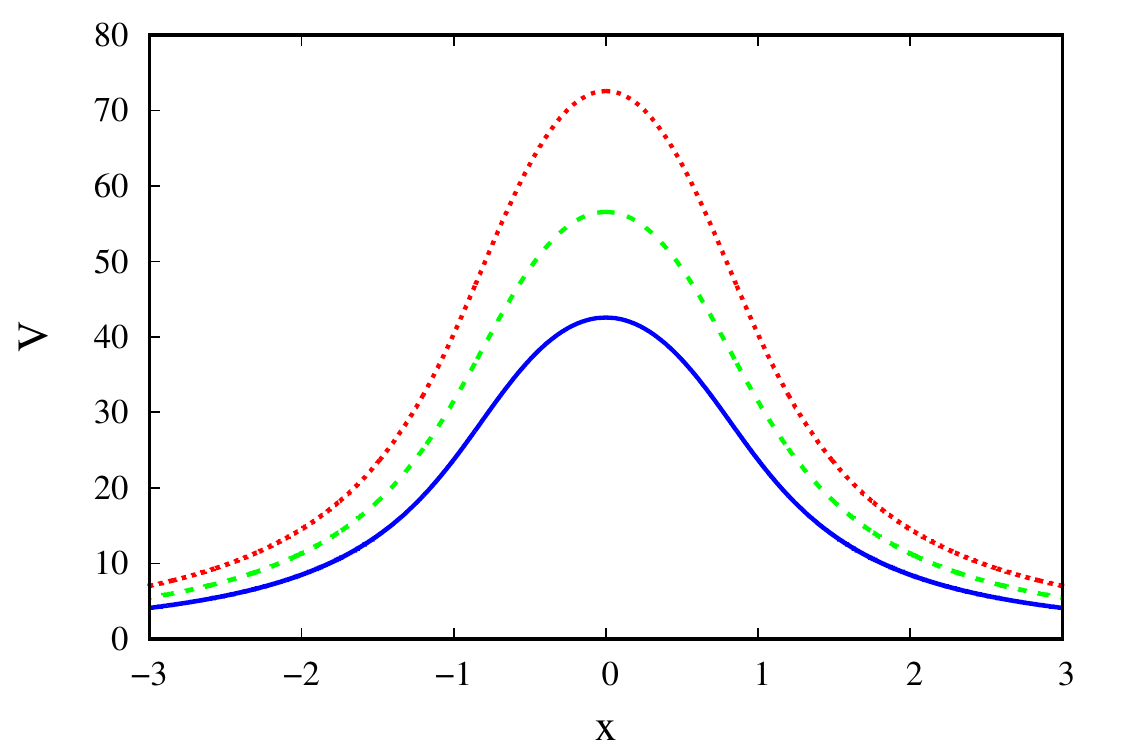}  \
\includegraphics[width=0.3\textwidth]{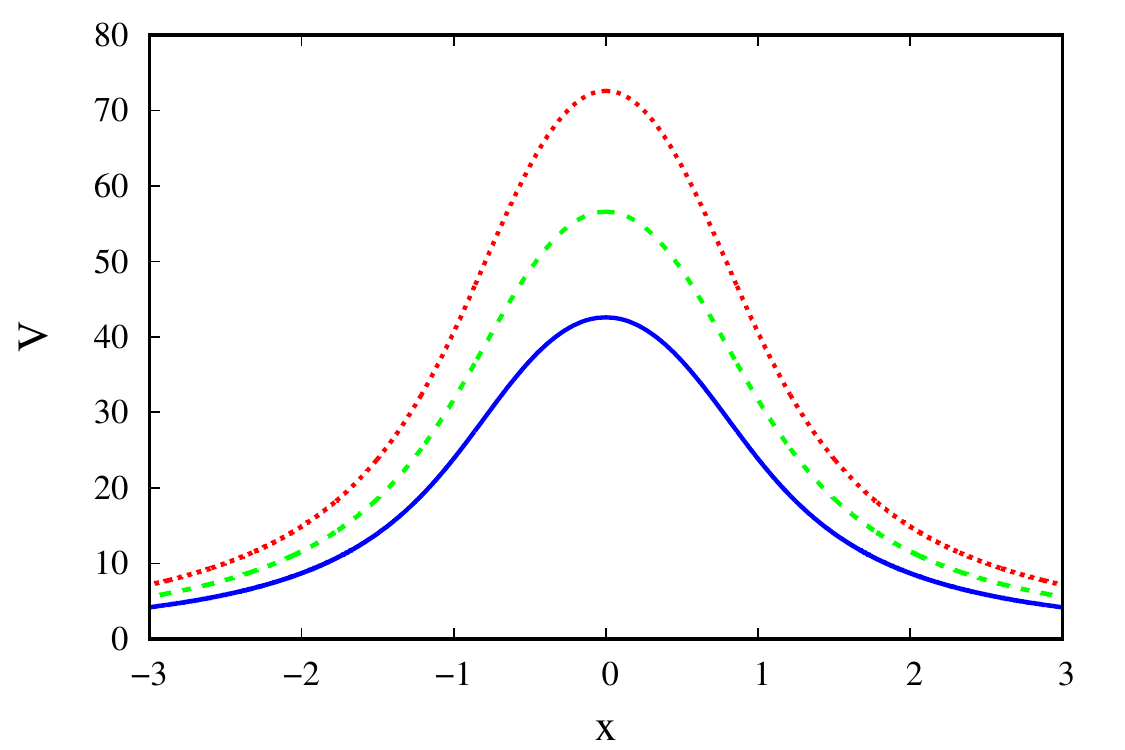}  \
\includegraphics[width=0.3\textwidth]{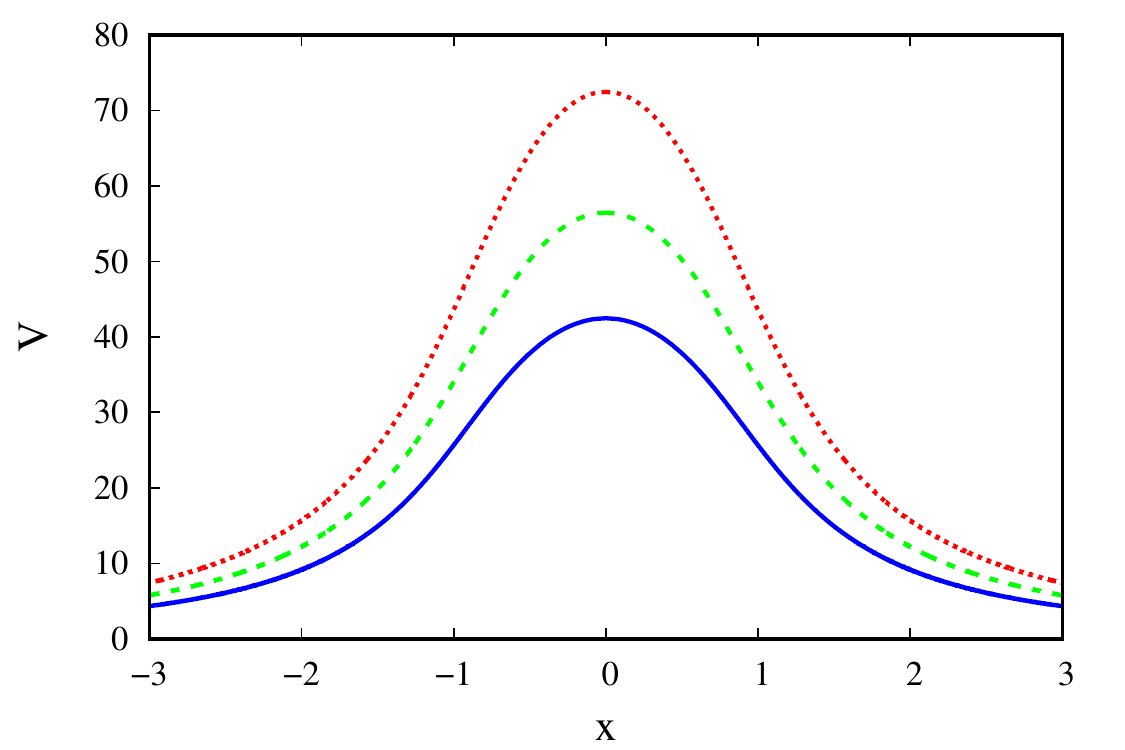}  \
\includegraphics[width=0.3\textwidth]{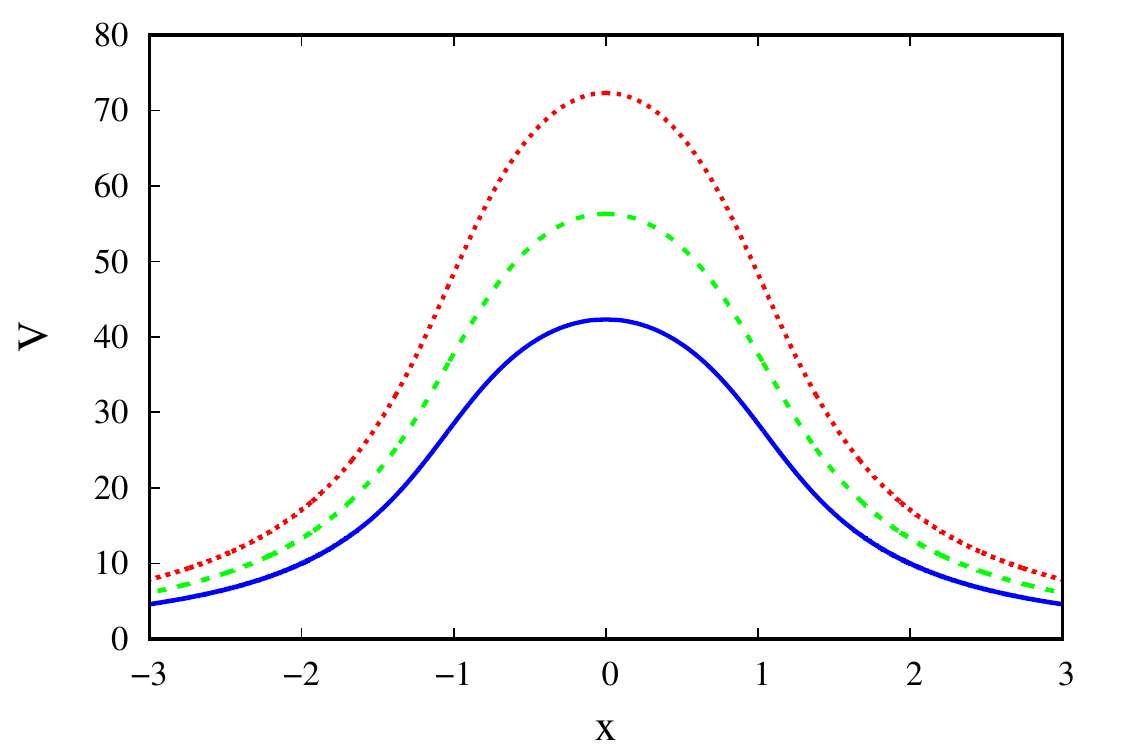}  \
\includegraphics[width=0.3\textwidth]{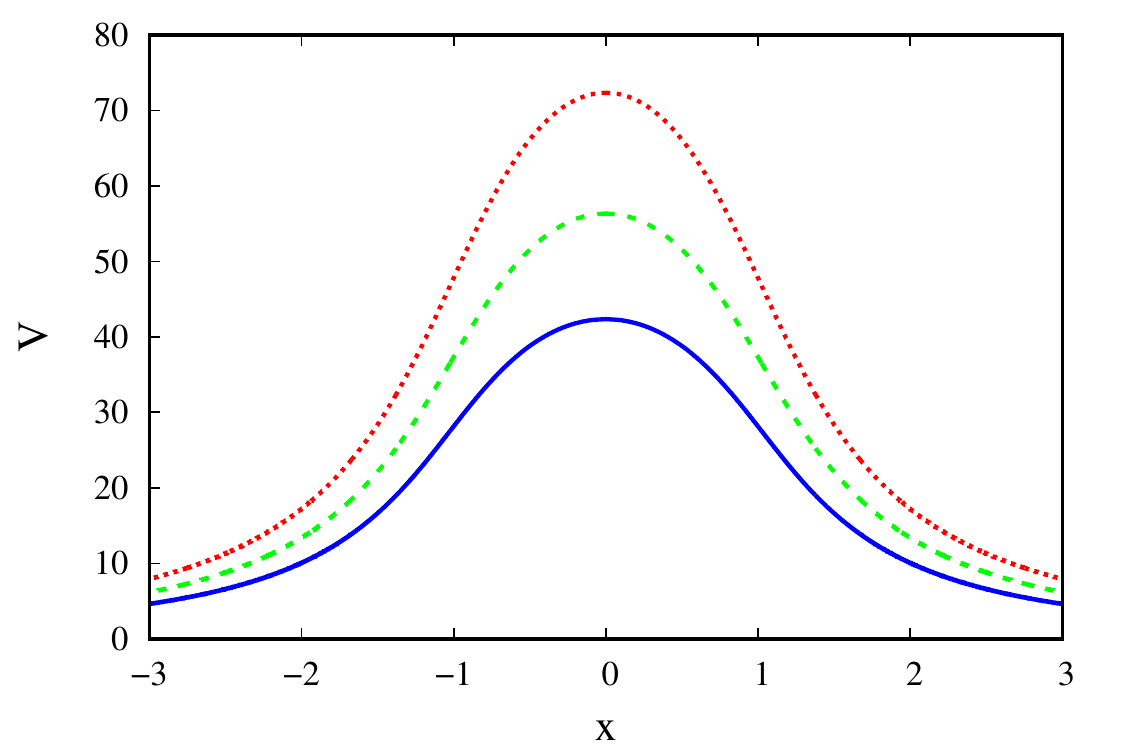}  \
\includegraphics[width=0.3\textwidth]{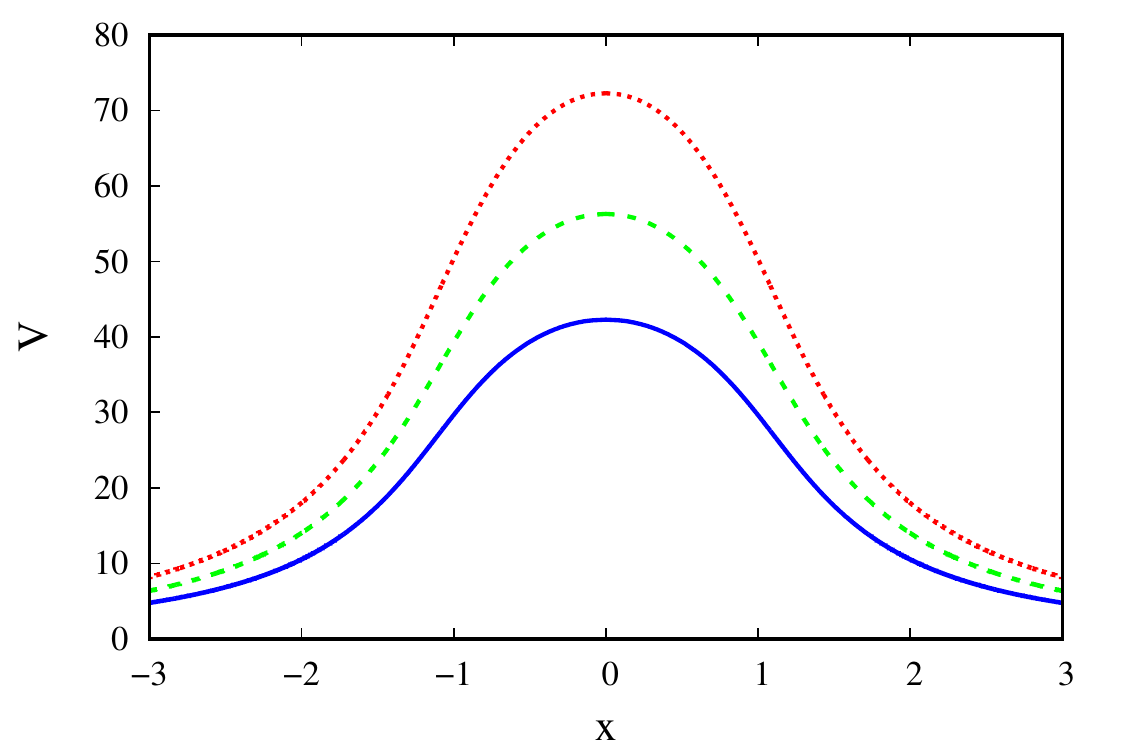}  \
\includegraphics[width=0.3\textwidth]{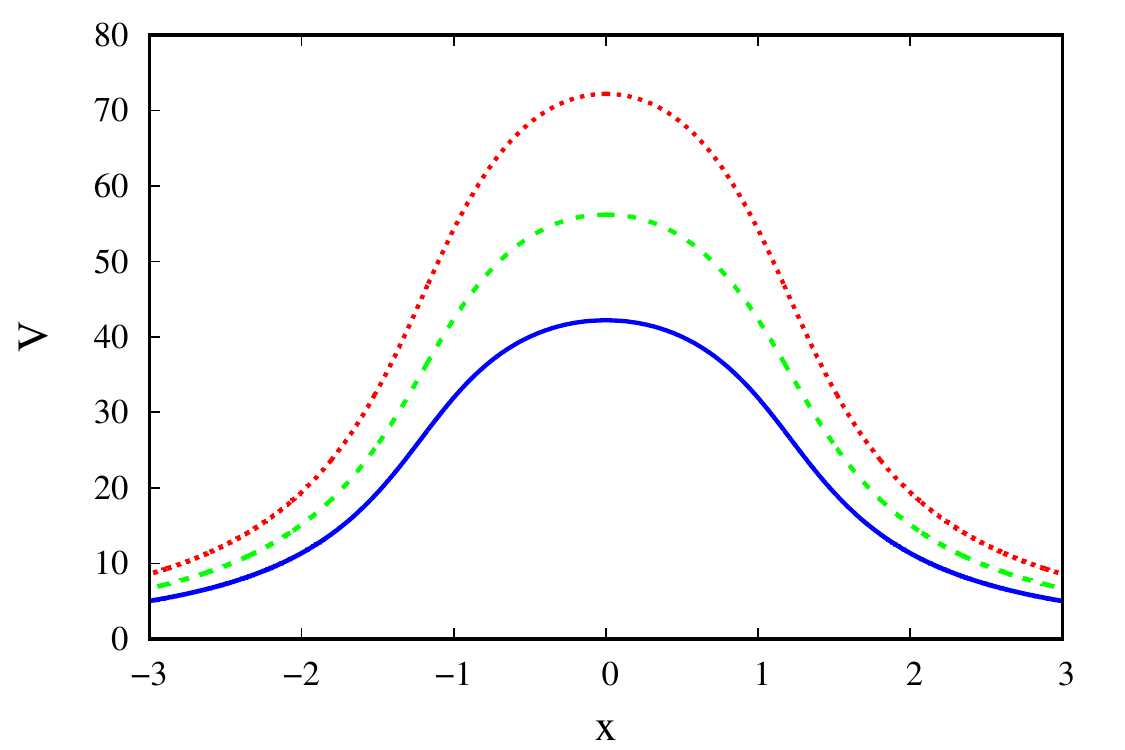}  
\caption{\label{potential}
The potential of perturbation as a function of $x$ (the tortoise coordinate) for $l = 6$ (blue), $l = 7$ (green), $l = 8$ (red).  We have set $c = 0.4$ (first row),
$c=0.8$ (second row), $c=1.2$ (third row) and $c=1.6$ (fourth row). For each row we have $d = 0.2$ (left panel), $d = 0.6$ (center panel) and $d = 1.4$ (right panel).}
\end{figure*}

In what follows, we shall show numerical results for the QNM associated to scalar perturbations of the model. As we stated previously, the implementation of the WKB method requires a bell--shaped potential as a function of the tortoise coordinate as shown in Fig. (\ref{potential}) for different values of the parameters involved. We note that the peak of the potential decreases as $l$ decreases. Besides, the potential spread out and its peak decreases as $c$ grows.

\begin{figure*}[hbt!]
\centering
\includegraphics[width=0.3\textwidth]{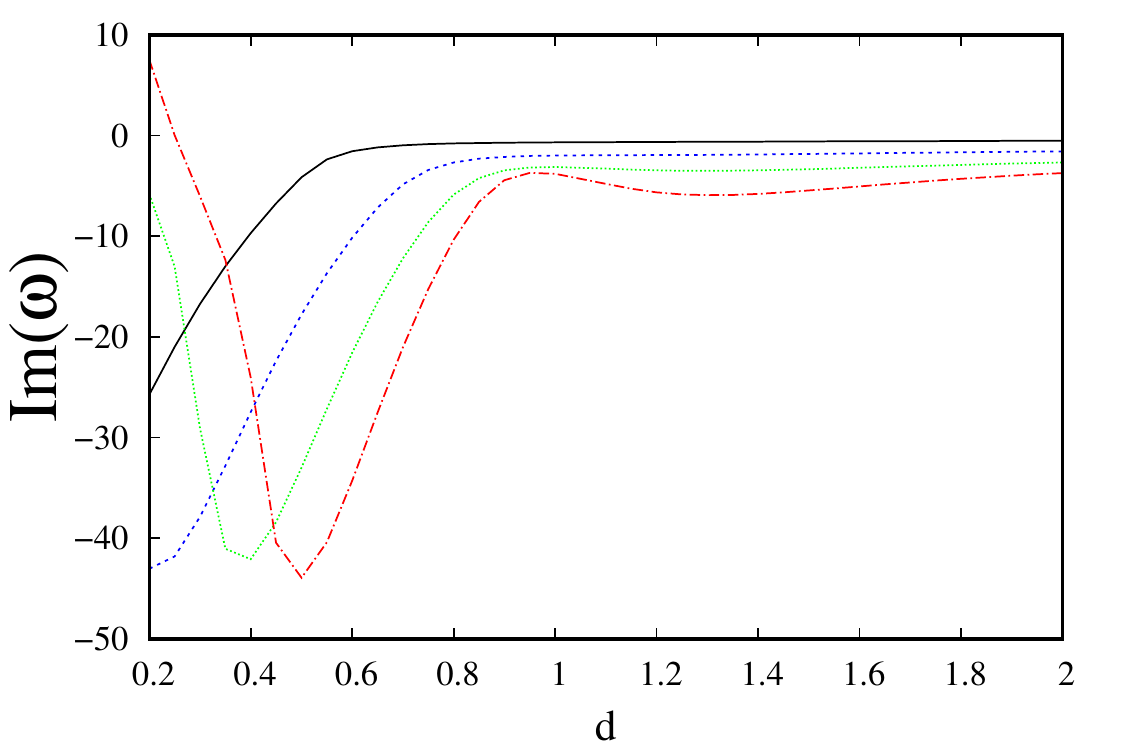}  \
\includegraphics[width=0.3\textwidth]{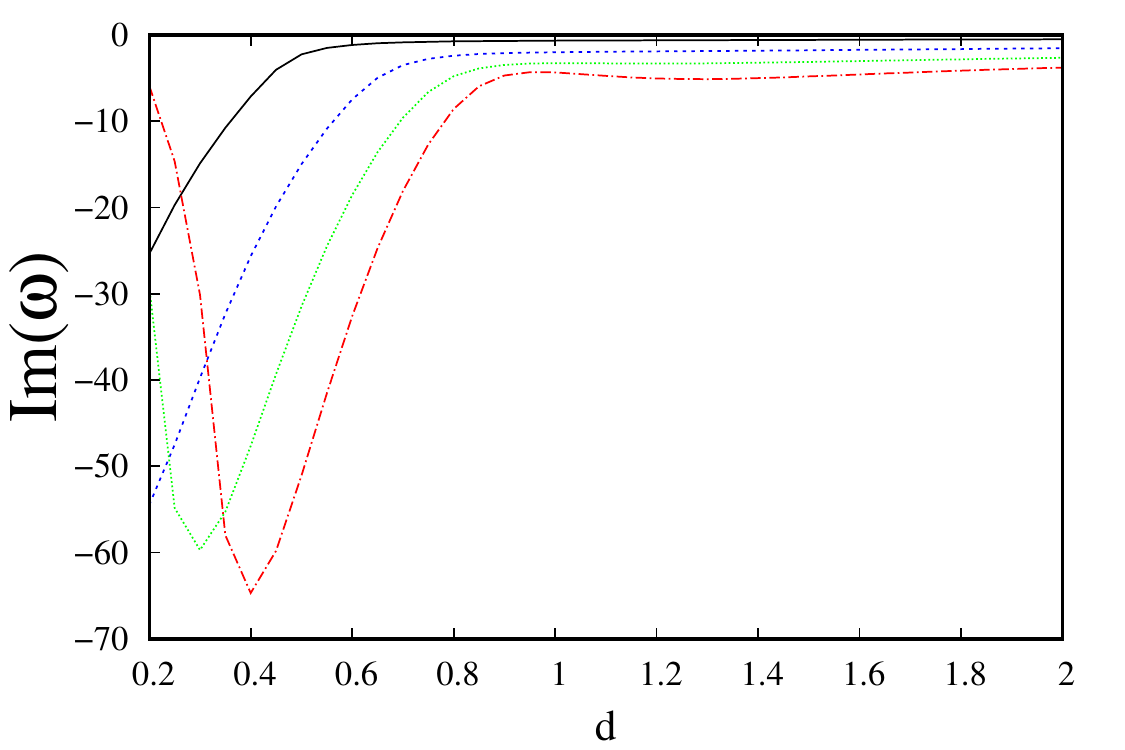}  \
\includegraphics[width=0.3\textwidth]{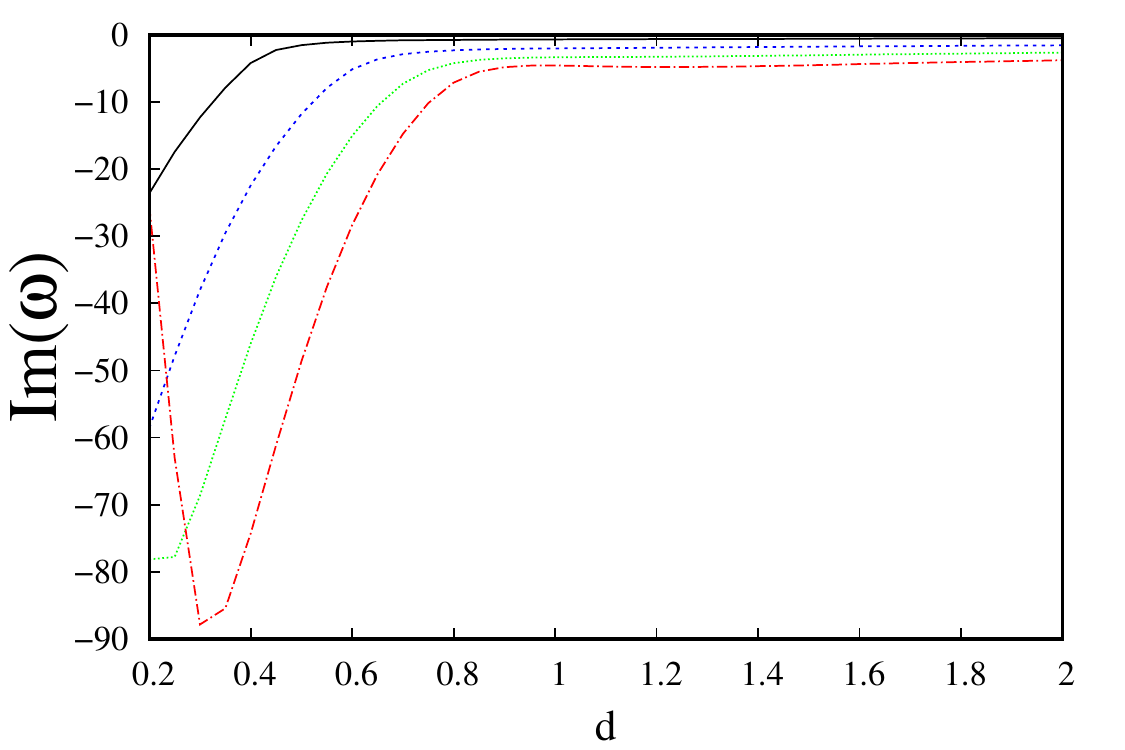}  
\includegraphics[width=0.3\textwidth]{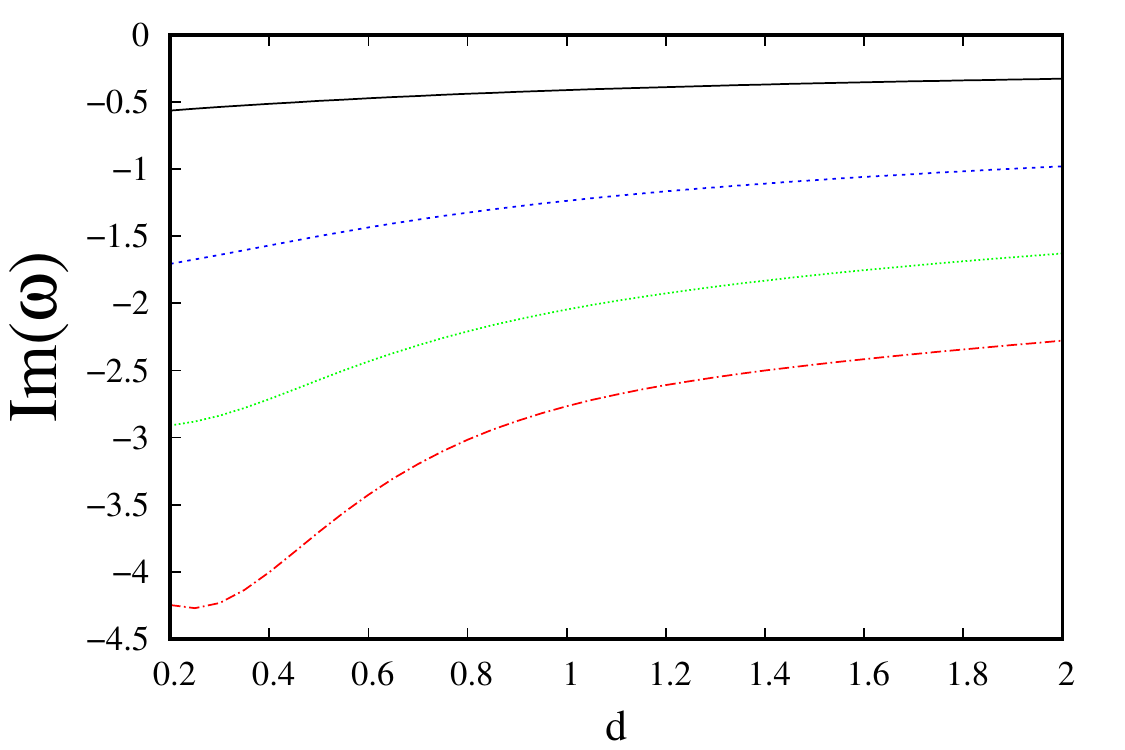}  \
\includegraphics[width=0.3\textwidth]{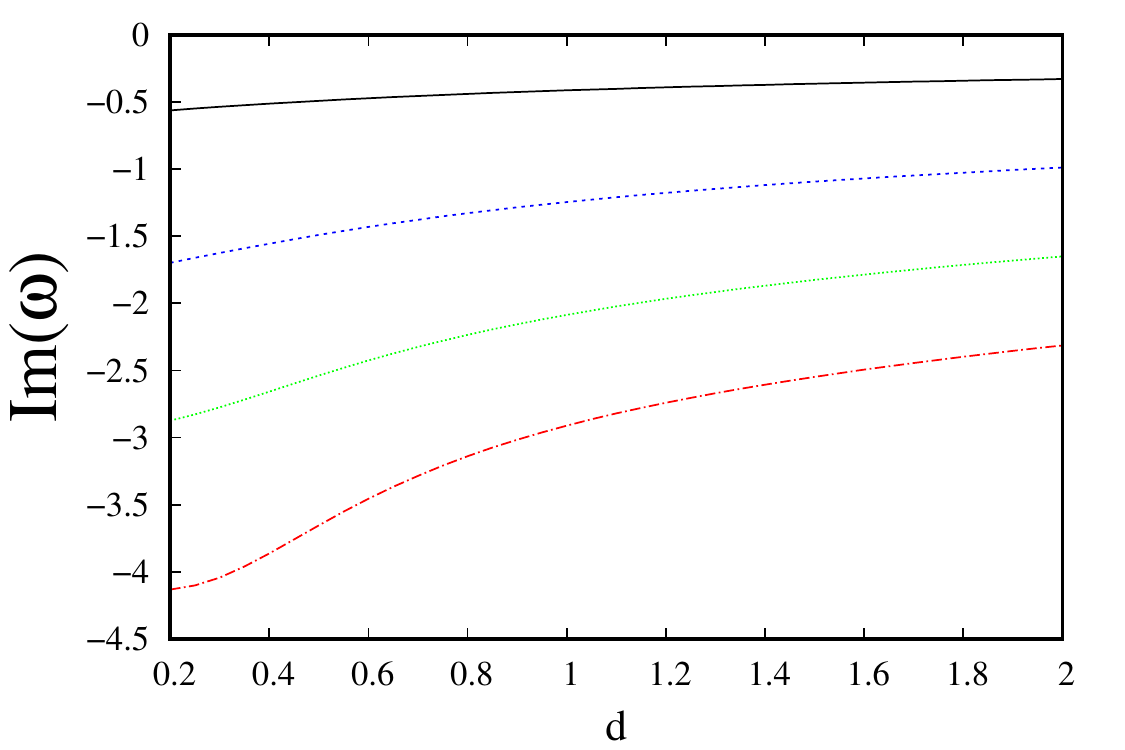}  \
\includegraphics[width=0.3\textwidth]{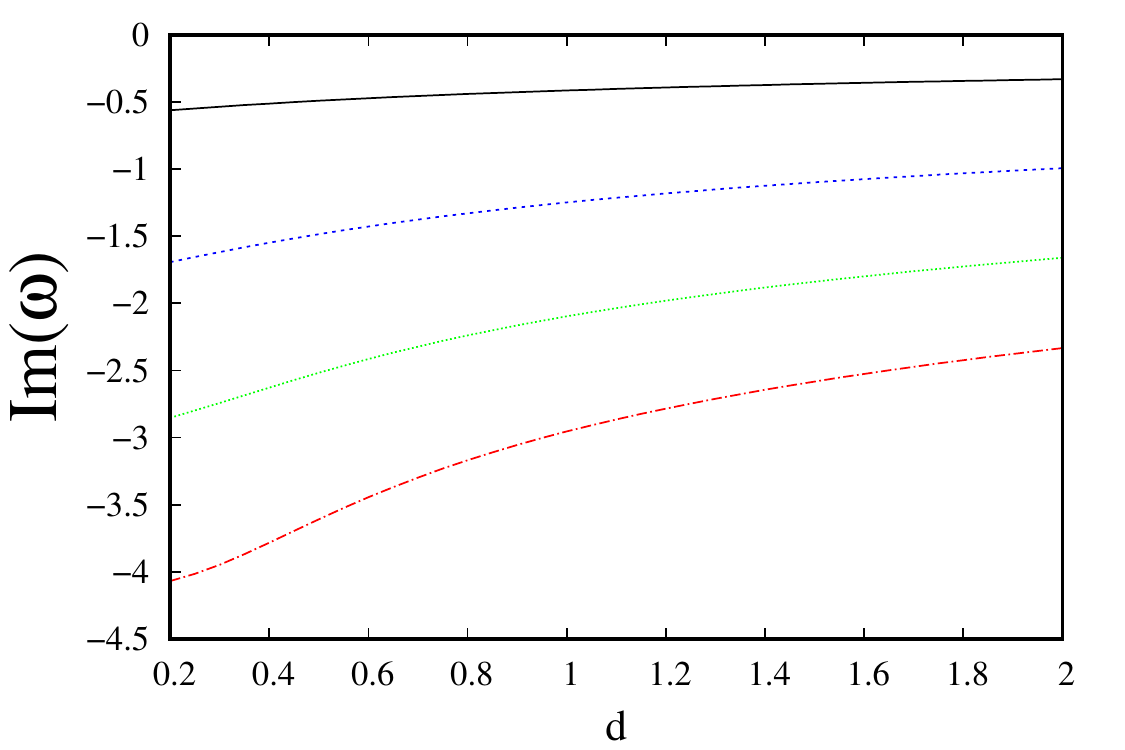}  \
\includegraphics[width=0.3\textwidth]{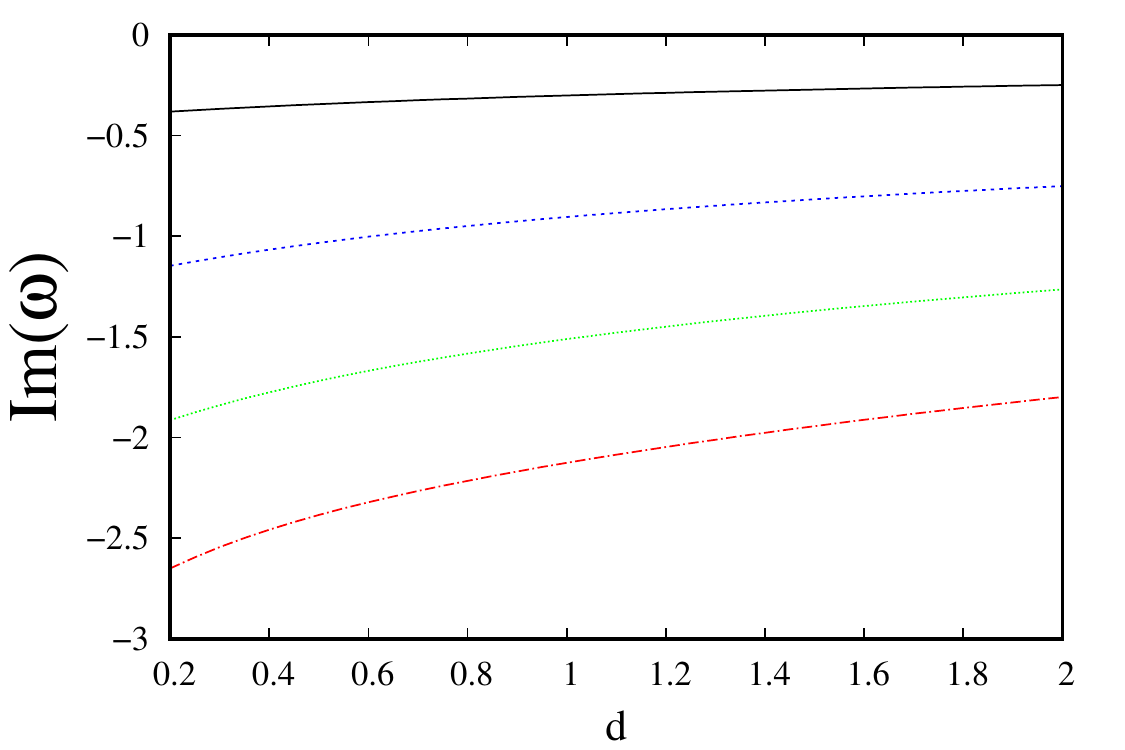}  \
\includegraphics[width=0.3\textwidth]{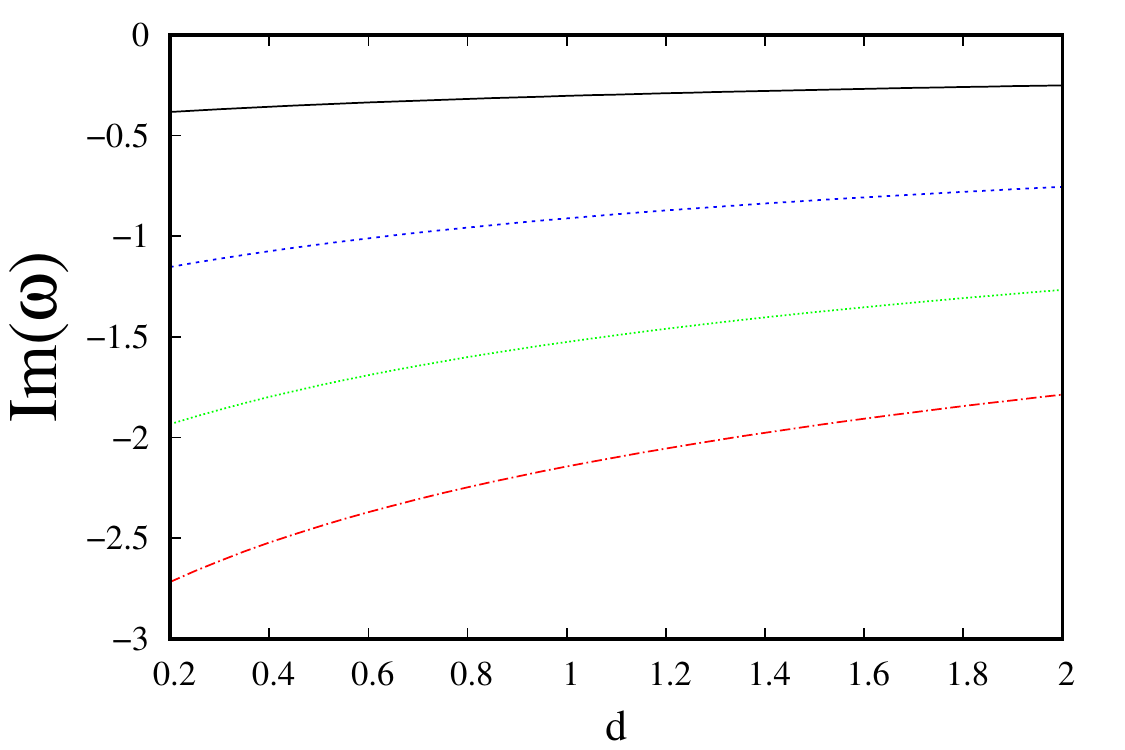}  \
\includegraphics[width=0.3\textwidth]{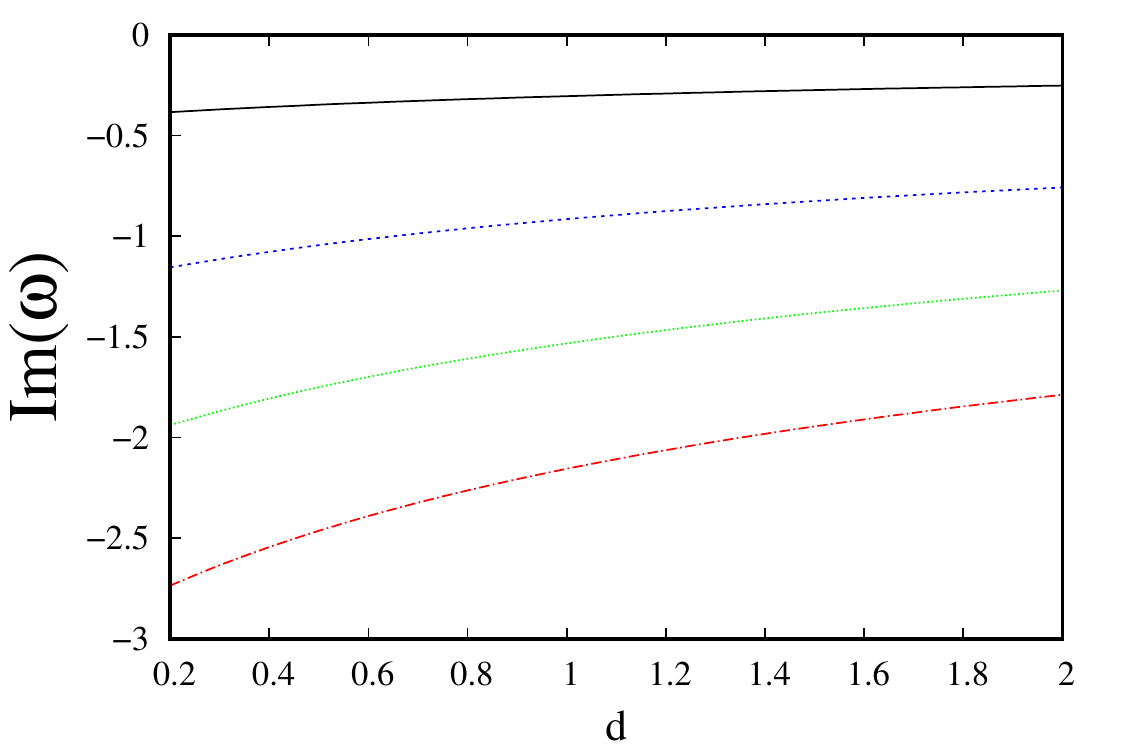}  \
\includegraphics[width=0.3\textwidth]{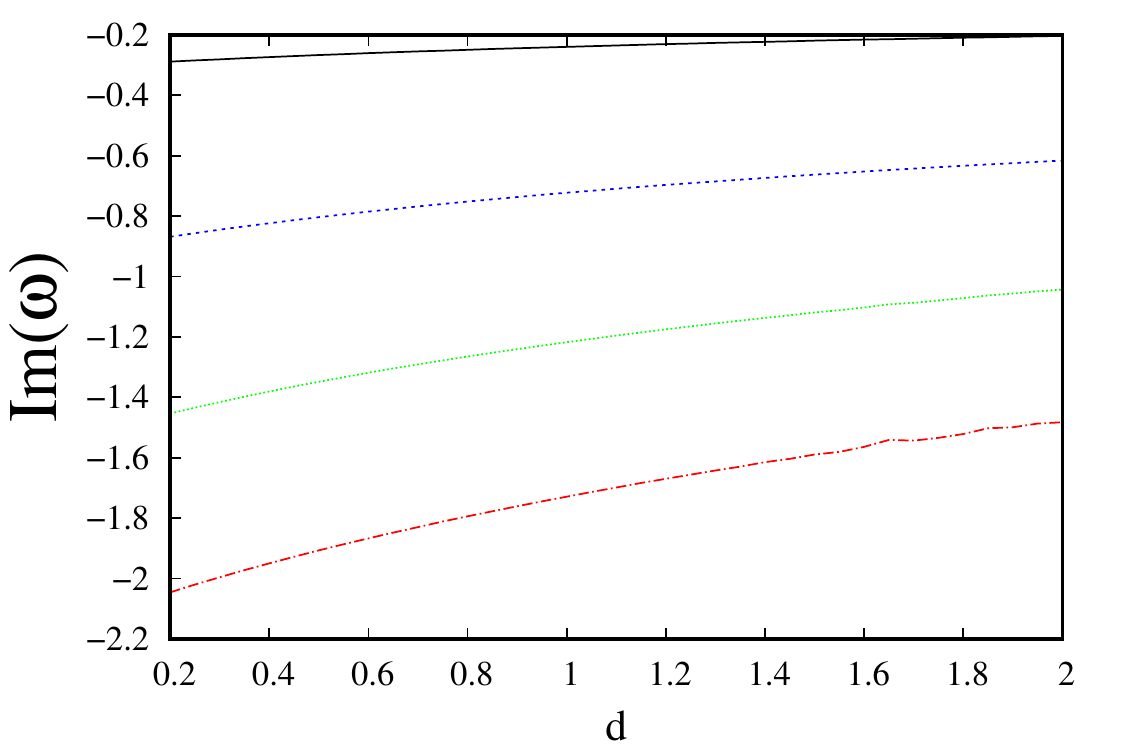}  \
\includegraphics[width=0.3\textwidth]{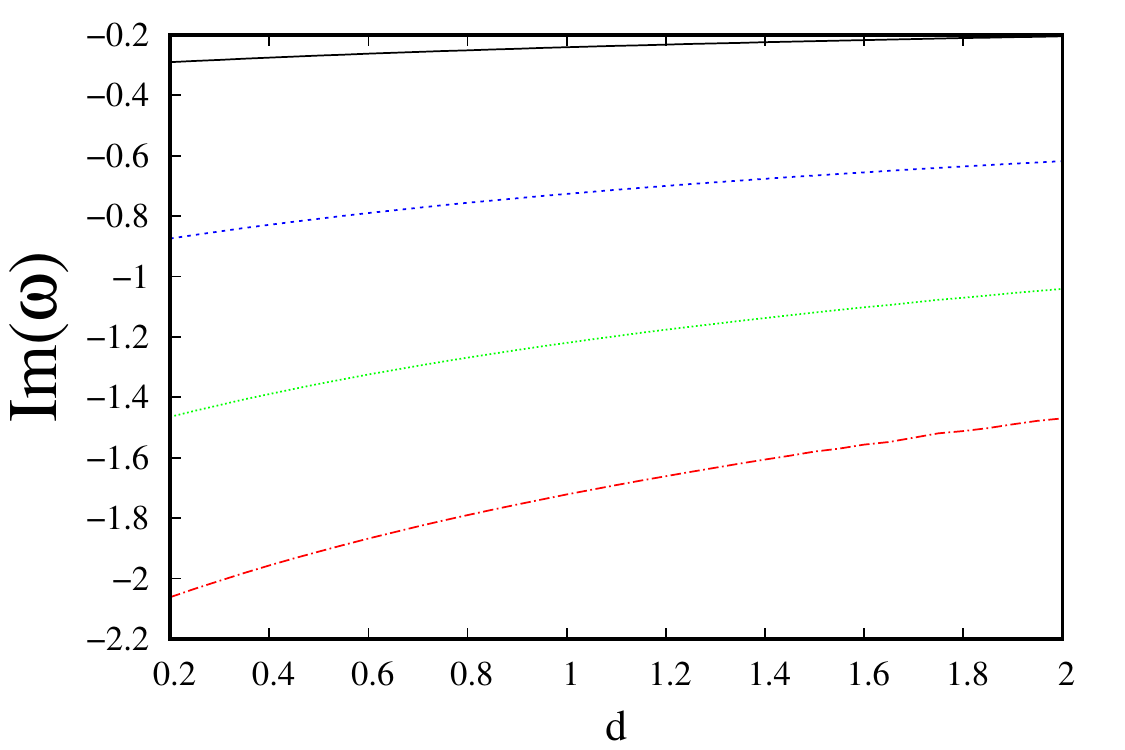}  \
\includegraphics[width=0.3\textwidth]{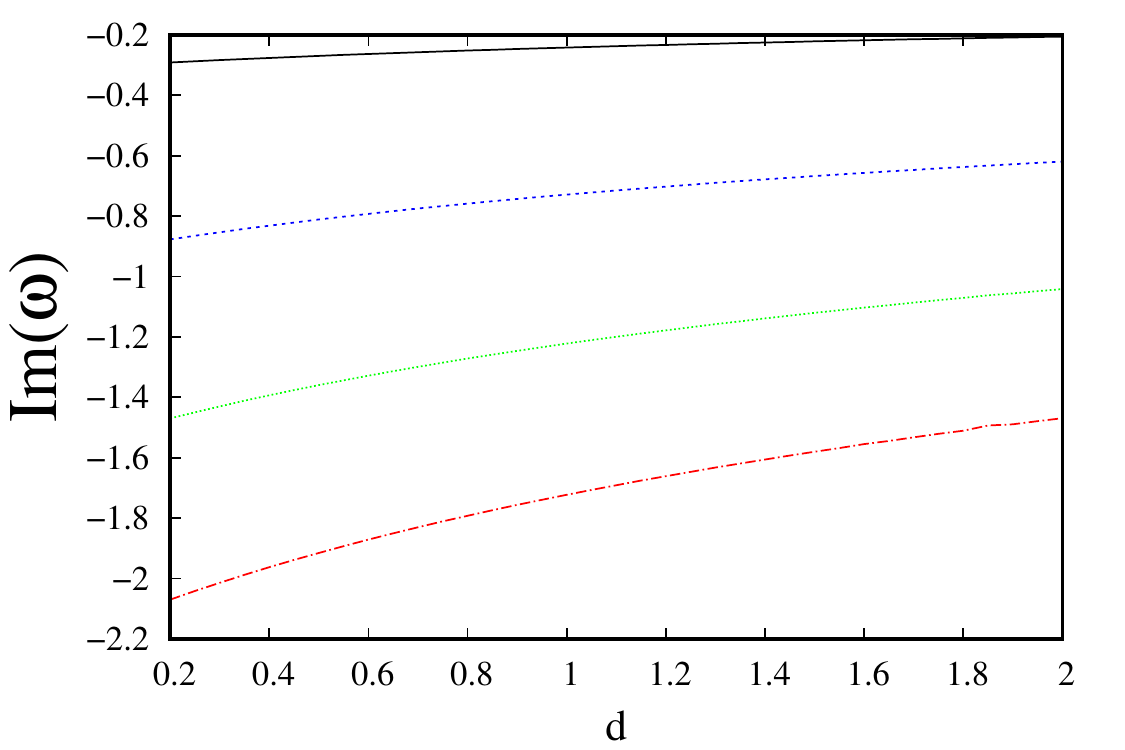}  
\caption{\label{QNMimg}
Imaginary part of the frequency as a function of $d$ for $n = 0$ (black), $n = 1$ (blue), $n = 2$ (green), $n = 3$ (red).  We have set $c = 0.4$ (first row),
$c=0.8$ (second row), $c=1.2$ (third row) and $c=1.6$ (fourth row). For each row we have $l = 6$ (left panel), $l = 7$ (center panel) and $l = 8$ (right panel).}
\end{figure*}

In Fig. \ref{QNMimg} we show the imaginary part of the frequency as a function of the parameter $d$. For $c=0.4$ (first row), we note that the profile is monotonously increasing and approaches asymptotically to zero. In contrast, for $n=2$ and $n=3$, $Im(\omega)$ decreases with $d$, reach a minimum and grows again approaching asymptotically to zero. Particularly, for $l=6$, the frequency has positives values for $d\in(0.2,0.3)$ which means that the wormhole is unstable for this 
interval. For  $c=0.8$ (second row), $c=1.2$ (third row) and $c=1.6$ (fourth row), $Im(\omega)$ is an increasing function and is always negative for the values of $d$ under consideration so the solution can be considered as stable under scalar perturbation for this parameters. We also note that the value of $Im(\omega)$ decreases as the overtone $n$ increases.

\begin{figure*}[hbt!] 
\centering
\includegraphics[width=0.3\textwidth]{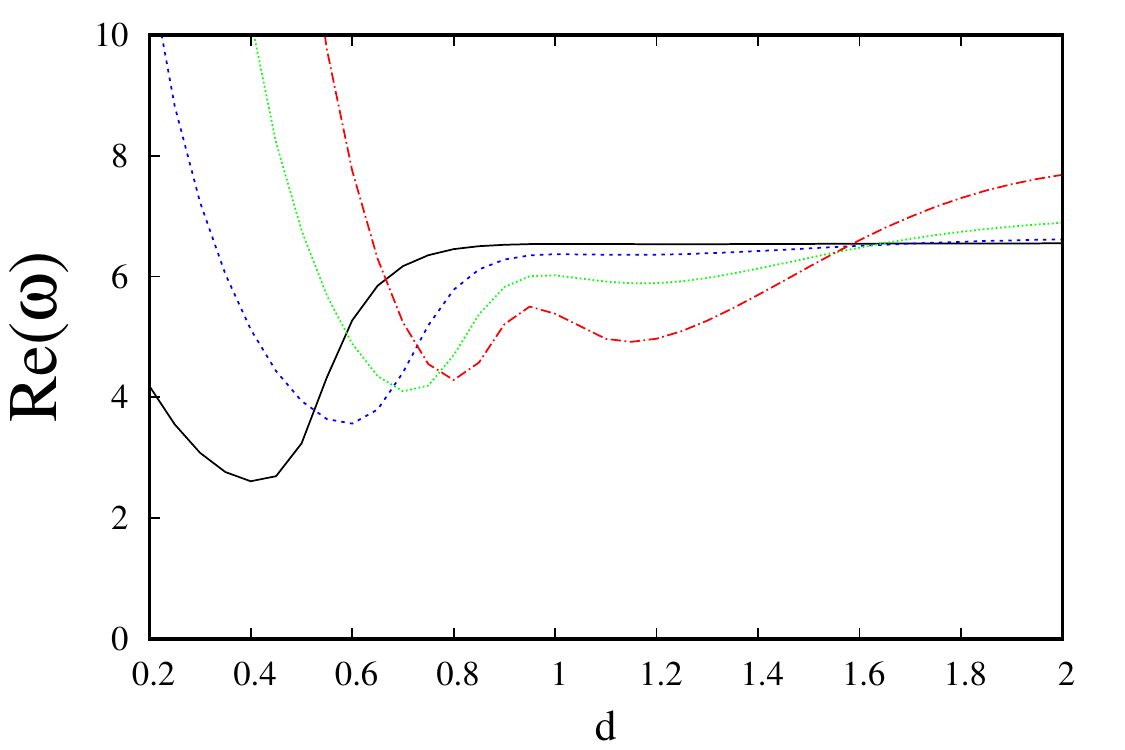}  \
\includegraphics[width=0.3\textwidth]{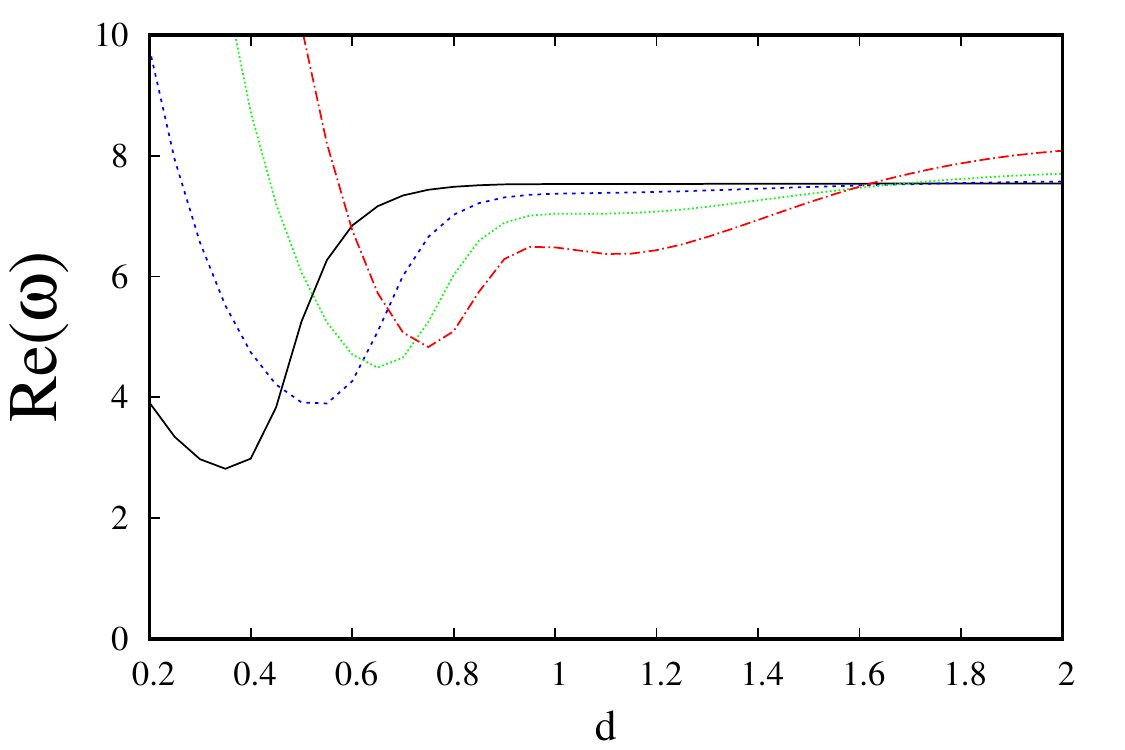}  \
\includegraphics[width=0.3\textwidth]{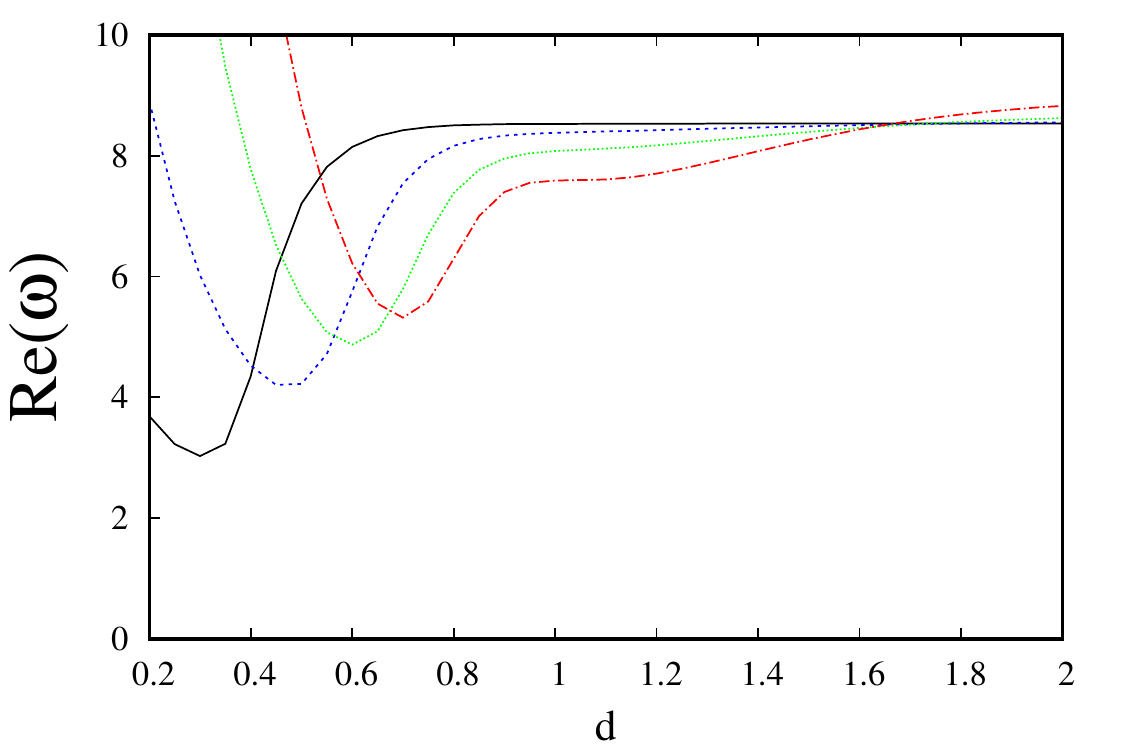}  
\includegraphics[width=0.3\textwidth]{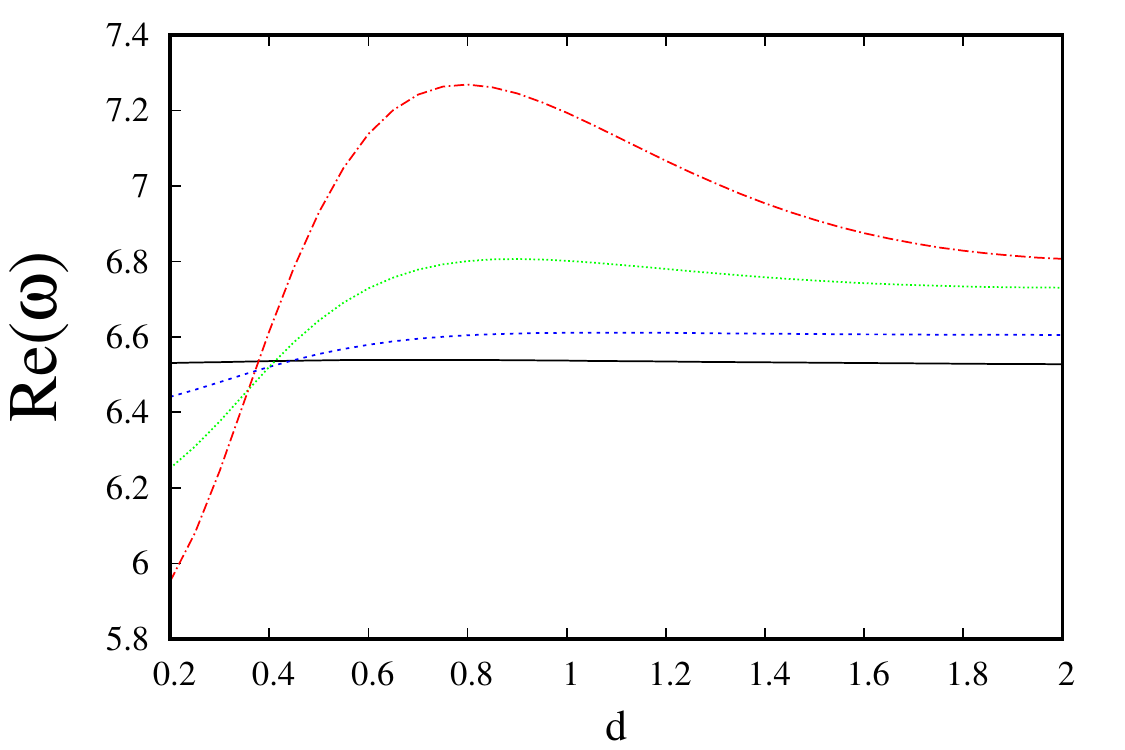}  \
\includegraphics[width=0.3\textwidth]{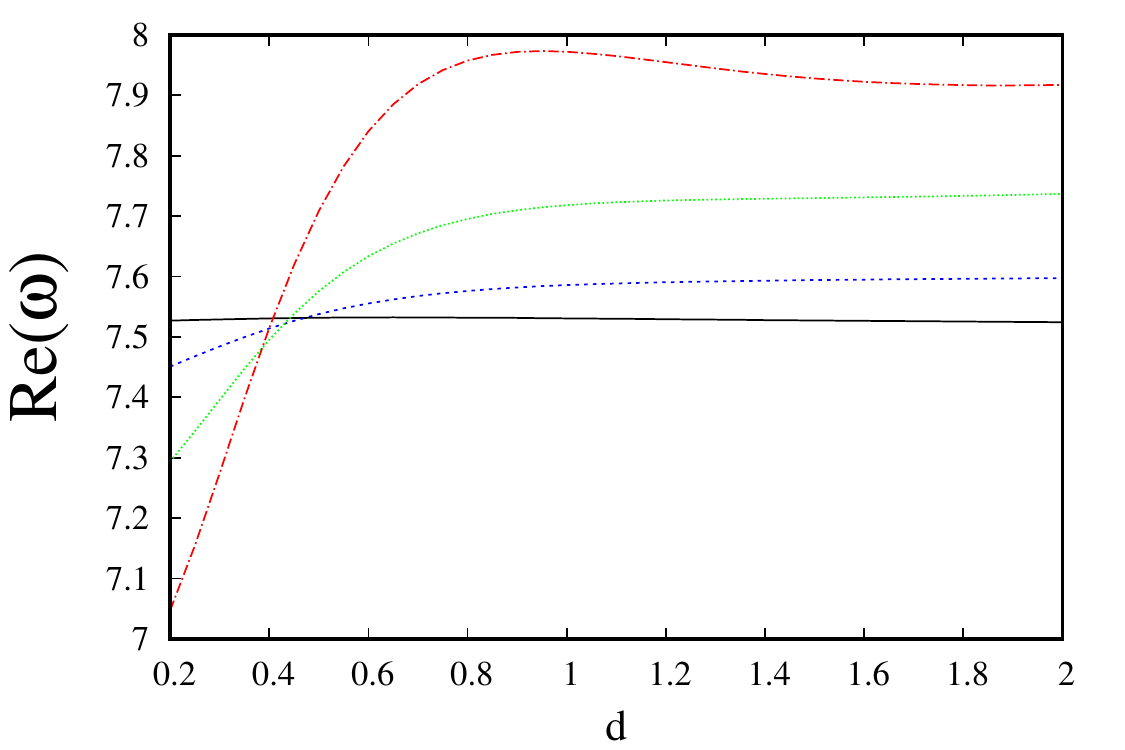}  \
\includegraphics[width=0.3\textwidth]{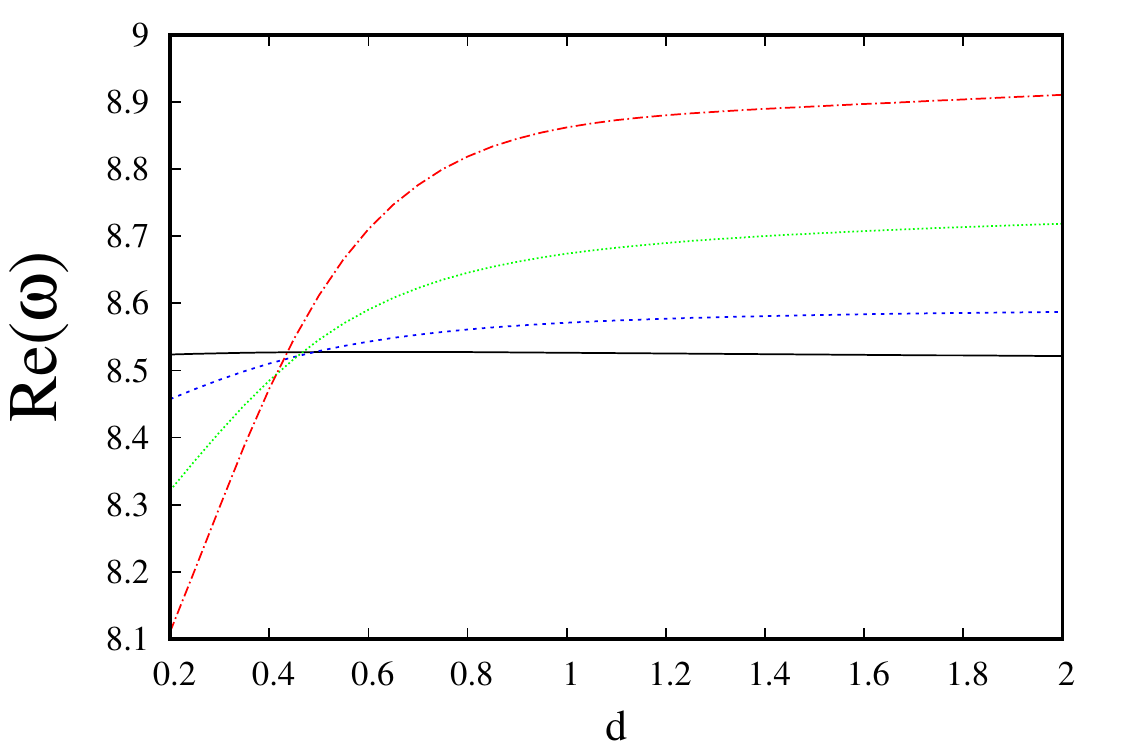}  \
\includegraphics[width=0.3\textwidth]{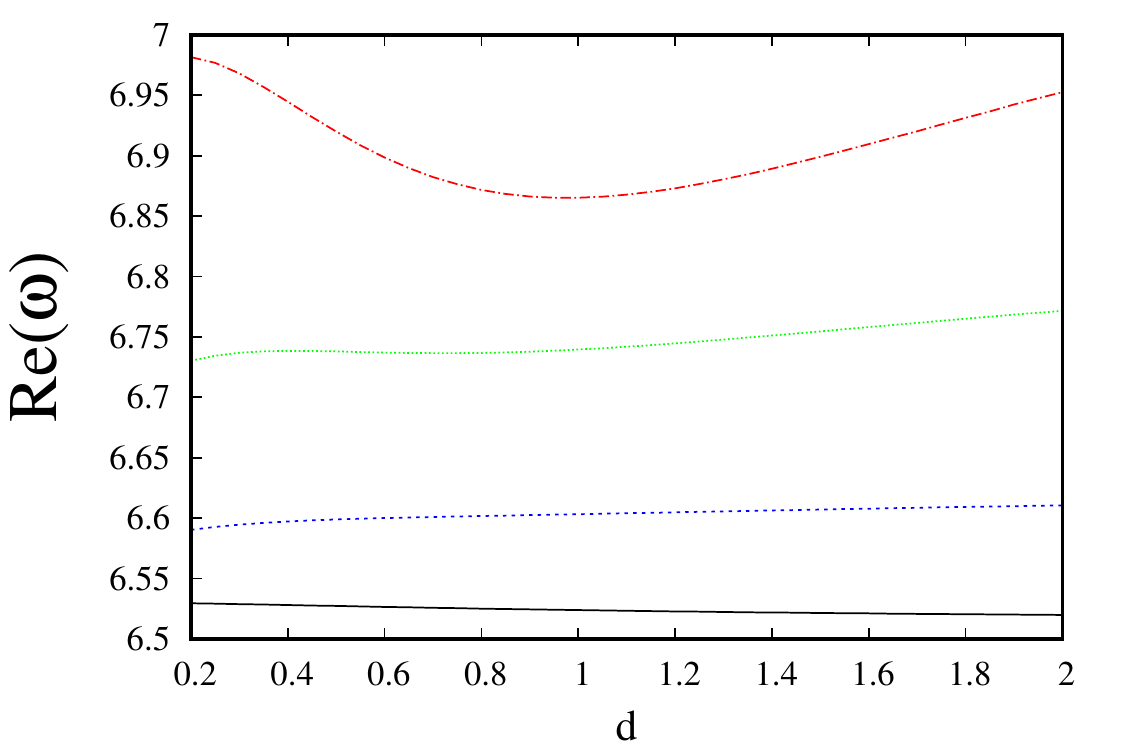}  \
\includegraphics[width=0.3\textwidth]{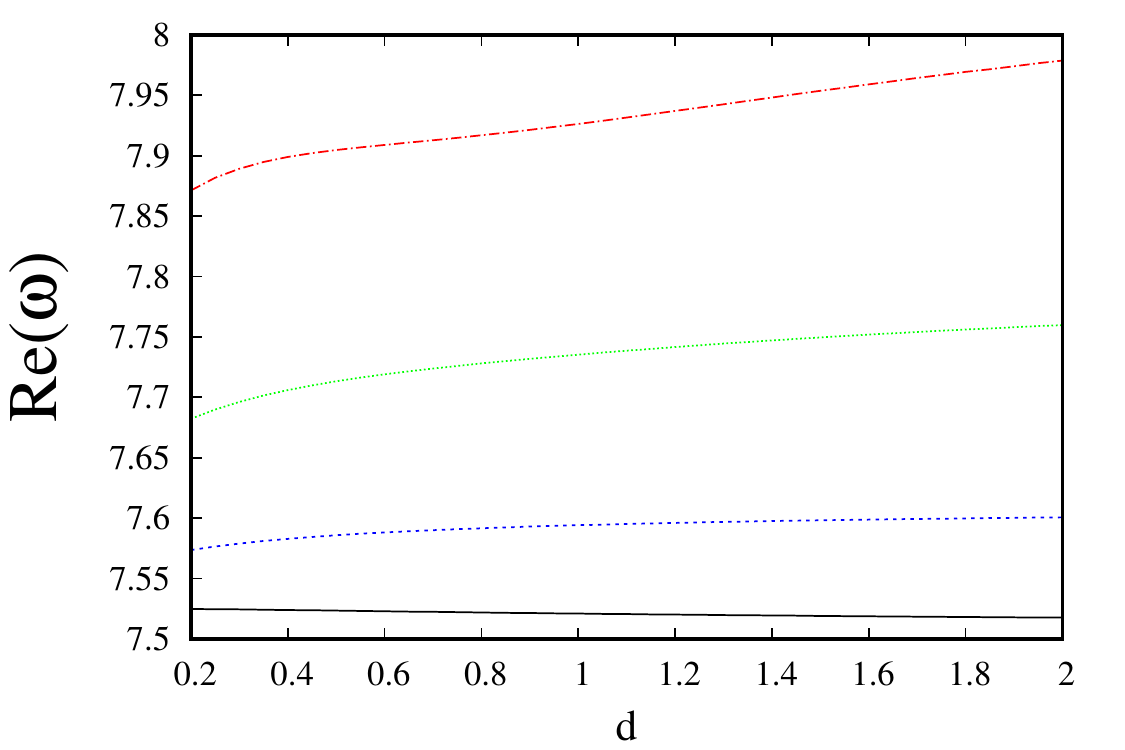}  \
\includegraphics[width=0.3\textwidth]{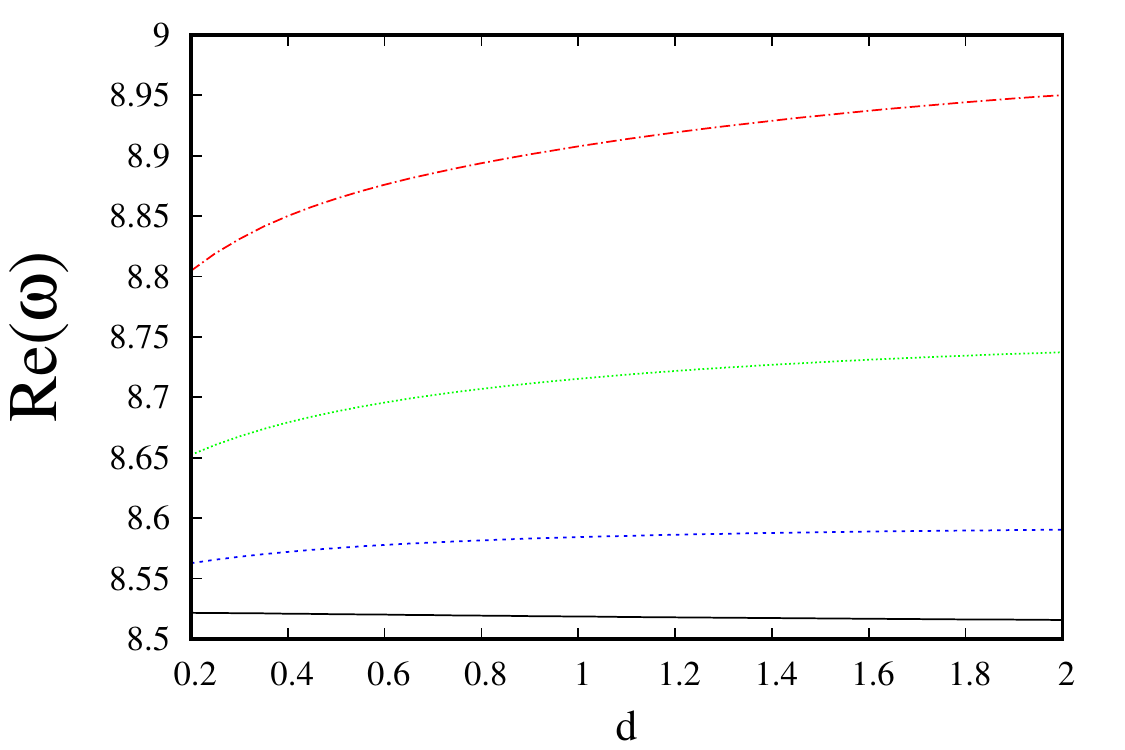}  \
\includegraphics[width=0.3\textwidth]{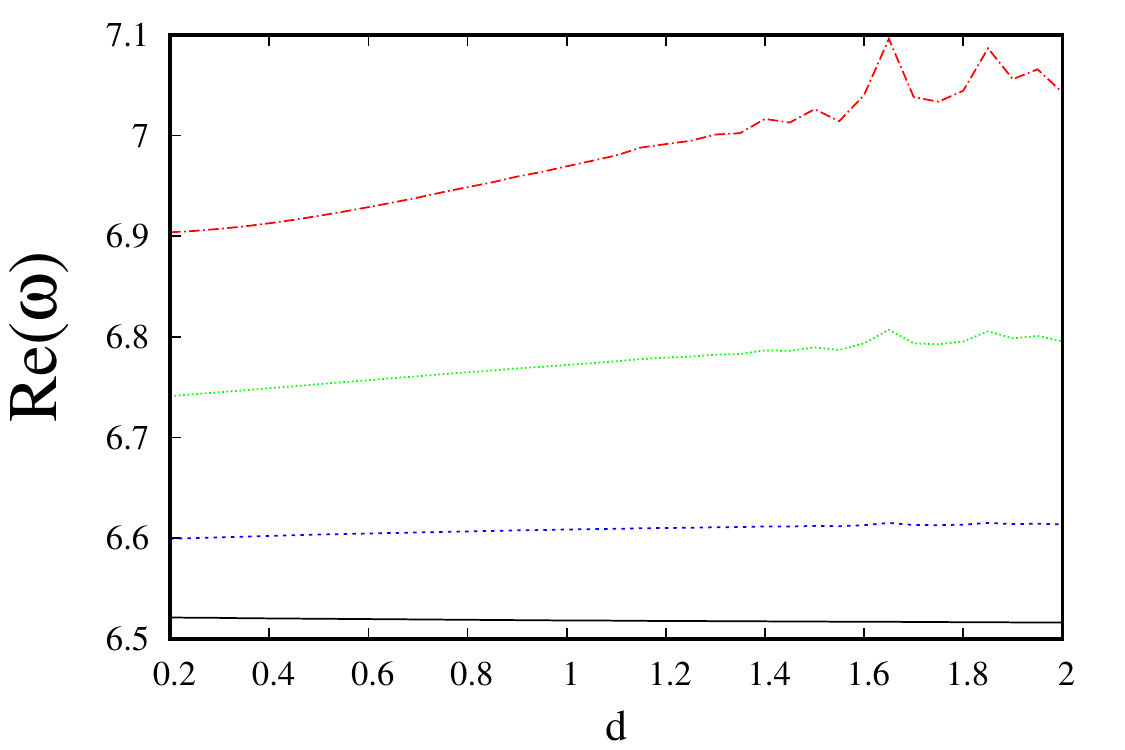}  \
\includegraphics[width=0.3\textwidth]{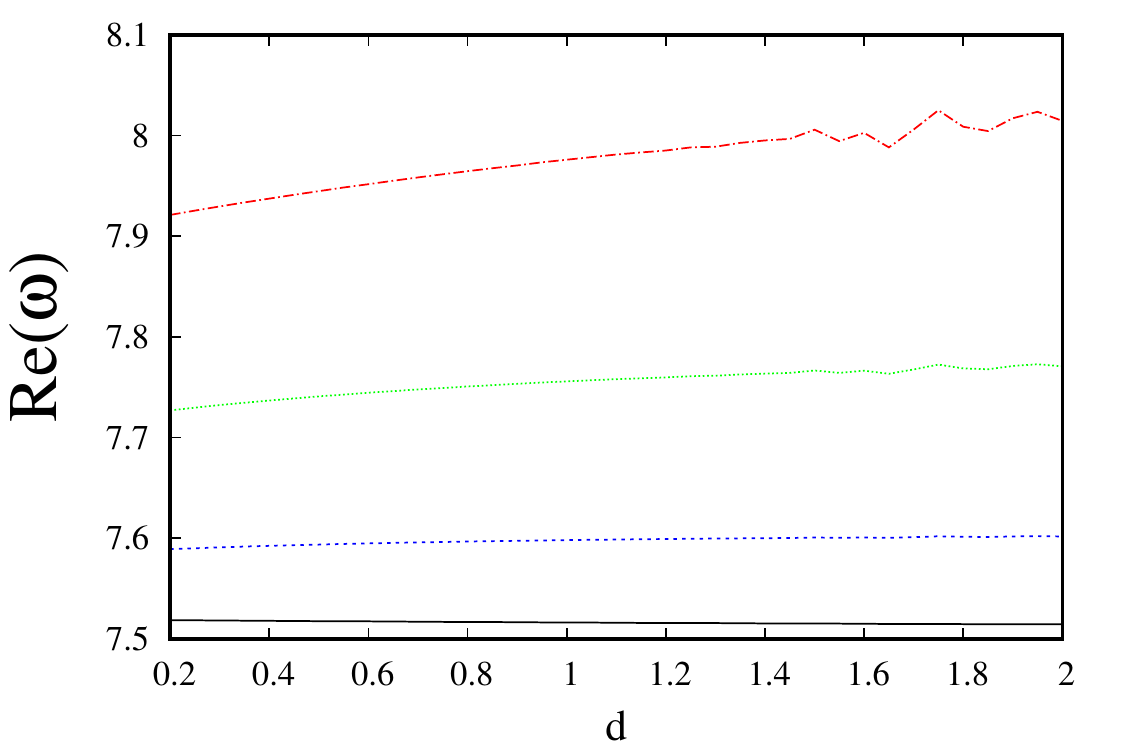}  \
\includegraphics[width=0.3\textwidth]{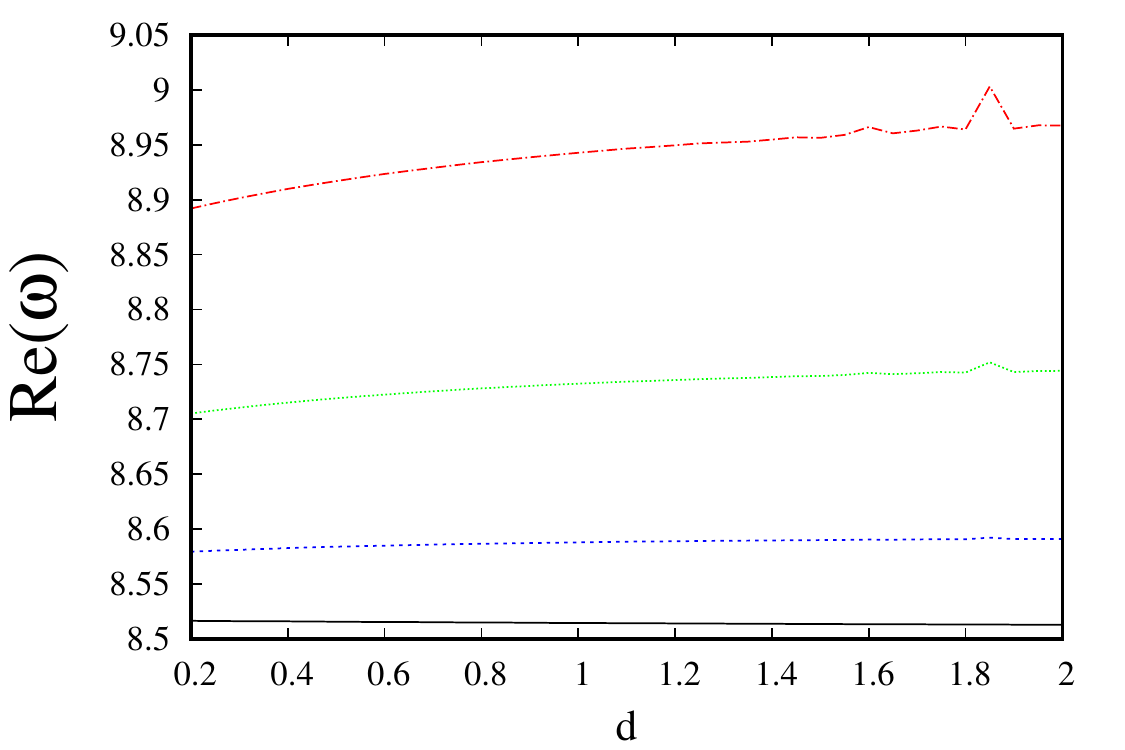}  
\caption{\label{QNMreal}
Real part of the frequency as a function of $d$ for $n = 0$ (black), $n = 1$ (blue), $n = 2$ (green), $n = 3$ (red).  We have set $c = 0.4$ (first row),
$c=0.8$ (second row), $c=1.2$ (third row) and $c=1.6$ (fourth row). For each row we have $l = 6$ (left panel), $l = 7$ (center panel) and $l = 8$ (right panel).}
\end{figure*}

In Fig. \ref{QNMreal} we show the $Re(\omega)$ as a function of $d$ for different values of the parameter $c$. For $c=0.4$ (first row), the profile  reaches a minimum located at different values of $d$ depending of the overtone. More precisely, as $n$ increases, the location of the minimum shift to larger values of $d$. Besides, except for $n=3$, the $Re(\omega)$ converge to the same value as $d$ grows which means that the oscillatory behaviour of the signal in indistinguishable for each overtone for large $d$. The asymptotic line shift to bigger values of $Re(\omega)$ as $l$ increases. For $c=0.8$ (second row) except for $n=0$ which remains constant, the $Re(\omega)$ reaches a maximum in contrast to the previous case. Moreover, the signal approaches asymptotically to a constant value which is different for each overtone. For $c=1.20$ (third row) and $l=6$, the signal is constant for $n=0$ and increases monotonously for $n=1$ and $n=2$ and reach a minimum for $n=3$. In contrast to the previous cases, the $Re(\omega)$ do not approach asymptotically to any value in the interval under consideration. For $l=7$ and $l=8$, except for the lowest overtone, $Re(\omega)$ increases monotonously. For $c=1.6$, the signal is constant for $n=0$, and increases monotonously for $n=1$ and $n=2$. For $n=3$, the signal increases with but undergoes some oscillatory behaviour for large $d$. This behaviour can be associated to numerical instabilities. 

Based on the previous results, at this point some comments are in order. First, for $c=0.4$, the behaviour of signal for large $d$ is almost independent of the overtone. In particular, the damping factor, given by $e^{Im(\omega)}$, is almost the same for each $n$ under consideration. Similarly, the oscillatory behaviour is almost monochromatic: $Re(\omega)$ converge to the same value as $l$ grows. In this regard, if the main goal is to differentiate the behaviour from overtones for $c=0.4$, our model must be tested in the interval $d\in(0.2,1)$. It is worth mentioning that the model seems unstable in this interval for $n=3$. However, it is a well known fact that the model works well for lower overtones so that such instability could be associated to inaccuracy of the method. Second, for $c=0.8$ the behaviour clearly depends on the value of the overtone in the whole interval of $d$ under consideration. In particular, the damping is both stronger for large $n$ and weaker a $d$ grows. Regarding the $Re(\omega)$, the behaviour of the oscillatory part depends on the value of $d$. Indeed, $Re(\omega)$ decreases as $n$ grows in $d\in(0.2,0.4)$ but increases for large $n$ in $d\in(0.4,2)$. Interestingly, all the frequencies of the oscillatory part coincide for $d=0.4$.  Finally, for $c=1.2$
and $c=1.6$, the pattern is clear: the damping is stronger as $n$ increases and $d$ decreases and the frequency of the oscillations grow as $n$ grows.

\section{Conclusions}
In this work we obtained a traversable wormhole 
with vanishing radial tidal force by proposing a general embedding function with some free parameters. The  parameters were reduced by imposing the basic requirements that must be satisfied by a wormhole geometry. In particular, we demanded the existence of a minimum radius (which defines the throat of the hole) and the flaring out condition. In order to explore how the geometry behaves in terms of the remaining parameter we analyzed both the quantifier of the exotic matter and the quasi normal modes of the solution. As a results, we observed that the solution requires a finite amount of exotic matter that decreases for certain values of the parameters involved. Besides, we obtained that in general the solution seems stable after scalar perturbations. Indeed, the imaginary part of the quasinormal frequencies remains negative which leads to a suitable damping factor for the signal.

\bibliography{references.bib}
\bibliographystyle{unsrt}
\end{document}